\documentclass[reprint,aps,prd,twocolumn,notitlepage,nofootinbib,showpacs,preprintnumbers,superscriptaddress]{revtex4-1}
\usepackage[T1]{fontenc}
\usepackage[latin9]{inputenc}
\usepackage{bm}
\usepackage{amsmath}
\usepackage{esint}
\usepackage{color}
\usepackage{graphicx}
\PassOptionsToPackage{normalem}{ulem}
\usepackage{ulem}

\usepackage{times}
\usepackage{hyperref}

\begin{document}

\title{Hamiltonian analysis of spatially covariant gravity}

\author{Xian Gao}%
    \email[Email: ]{gao@th.phys.titech.ac.jp}
    \affiliation{%
        Department of Physics, Tokyo Institute of Technology,\\ 
        2-12-1 Ookayama, Meguro, Tokyo 152-8551, Japan}

\date{\today}

\begin{abstract}
	We perform the Hamiltonian constraint analysis for a wide class of gravity theories that are invariant under spatial diffeomorphism.
	With very general setup, we show that different from the general relativity, the primary and secondary constraints associated with the lapse function $N$ become second class, as long as the lapse function $N$ enters the Hamiltonian nonlinearly. This fact implies that there are three degrees of freedom are propagating, of which two correspond to the usual tensor type transverse and traceless gravitons, and one is the scalar type graviton. By restoring the full spacetime diffeomorphism using the St\"{u}ckelberg trick, this type of spatially covariant gravity theories corresponds to a large class of single field scalar-tensor theories that possess higher order derivatives in the equations of motion, and thus is beyond the scope of the Horndeski theory.
\end{abstract}

\maketitle

\section{Introduction}

Models of inflation and dark energy, which attempt explaining the primordial and late time accelerating expansion of our universe, stimulate the study of theories beyond the simplest model based on general relativity (GR) with a cosmological constant.
One approach to these theories beyond GR (see \cite{Clifton:2011jh,Khoury:2013tda,Joyce:2014kja} for recent reviews) is to introduce degrees of freedom additional to the two tensor modes of GR.

These additional degrees of freedom are most straightforwardly realized by scalar fields.
Over the years, $k$-essence \cite{ArmendarizPicon:1999rj} was considered as the most general local theory for scalar field(s), which involves at most first derivatives of the field(s) in the Lagrangian.
Until recently, this understanding was systematically promoted to higher order in derivatives, by rediscovering the Horndeski theory \cite{Horndeski:1974wa} --- the most general covariant scalar-tensor theory involving up to second derivatives in the Lagrangian, while still leading to second order equations of motion for both scalar field and the metric --- as the ``generalized galileon'' \cite{Deffayet:2011gz}.
The ``second-order'' nature of Horndeski theory prevents it from extra ghost-like degrees of freedom and instabilities.
Lagrangian including second derivative of the scalar field firstly showed up in the decoupling limit of Dvali-Gabadadze-Porrati model \cite{Dvali:2000hr}, where the second derivative of the scalar field enters linearly in the Lagrangian.
Lagrangians for a single scalar field with nonlinear powers of second derivatives were systematically constructed in a Minkowski background  as the ``galileon'' model \cite{Nicolis:2008in}, which was then generalized to a curved background using the  ``covariantization'' procedure \cite{Deffayet:2009wt,Deffayet:2009mn}.
The ``generalized galileon'' \cite{Deffayet:2011gz} was constructed following the same procedure (see \cite{Deffayet:2013lga,Charmousis:2014mia} for reviews) and
was  shown to be exactly equivalent to the Horndeski theory \cite{Kobayashi:2011nu}.

Among the second derivatives in the Lagrangian, only those enter \emph{nonlinearly} will introduce higher order derivatives in the equations of motion, while those enter linearly are always safe.
The mathematical trick in Horndeski theory is to introduce nonlinear powers of second derivatives in terms of a special type of combinations, in which derivatives with the same spacetime index appear twice and only twice.
This fact implies that all the second derivatives with respect to a given spacetime index enter the Lagrangian linearly, although which is hidden in the polynomials nonlinear in the second derivatives. 
The covariant nature of the Horndeski theory implies that time and space are treated equally. However, one may slightly relax the above condition such that only second \emph{time} derivative appears linearly, while spatial derivatives can enter the Lagrangian in a rather arbitrary manner. This is the first hint that there might be more general theories beyond the Horndeski one.

On the other hand, new degrees of freedom may be introduced by reducing the symmetries of theories.
Therefore, an alternative approach to introducing additional degrees of freedom beyond GR, is to construct theories which do not respect the full diffeomorphism of GR.
A well studied example in this approach is the effective field theory (EFT) of inflation \cite{Creminelli:2006xe,Cheung:2007st} (which showed its first appearance in  ghost condensate \cite{ArkaniHamed:2003uy,ArkaniHamed:2003uz}), which describes the fluctuations around a time evolution homogeneous and isotropic background. 
Such a background explicitly breaks the full spacetime diffeomorphism of GR and thus a single scalar effective degree of freedom arises, which can be coupled to other matter fields \cite{Senatore:2010wk,Gwyn:2012mw,Noumi:2012vr,Ballesteros:2013nwa}.
This approach was further applied to the EFT of dark energy \cite{Creminelli:2008wc,Gubitosi:2012hu,Bloomfield:2012ff,Gleyzes:2013ooa,Bloomfield:2013efa}, where the dark energy can be described by a single scalar degree of freedom, which can also be coupled to several matter fields\footnote{See \cite{Tsujikawa:2014mba,Kase:2014cwa} for recent reviews on the EFT of inflation/dark energy as well as their relation with the Horndeski theory, and more references therein.} \cite{Gergely:2014rna,Kase:2014yya}.
Another extensively studied example, although initially motivated by a different purpose, is the Ho\v{r}ava gravity \cite{Horava:2009uw} and its healthy extension \cite{Blas:2009qj}, where a preferred foliation structure of spacetime was introduced.
In both the EFT of inflation/dark energy and the Ho\v{r}ava gravity, the full spacetime symmetry is spontaneously broken to the reduced spatial diffeomorphism on the spacelike hypersurfaces.
Therefore, we may refer to such kind of theories as ``spatially covariant gravity''. 
Remarkably, when fixing the gauge by choosing the scalar field as the time coordinate\footnote{This can be done as long as the gradient of the scalar field is kept as timelike, at least in a local region of spacetime.} (i.e., $\phi(t,\vec{x})=t$, which is often referred to as the ``unitary gauge''), the Horndeski theory \cite{Horndeski:1974wa,Deffayet:2011gz} can also be recast in terms of extrinsic and intrinsic curvatures, i.e. some type of the spatially covariant gravity \cite{Gleyzes:2013ooa}.
In \cite{Khoury:2011ay,Khoury:2013oqa}, gravity theories respecting only spatial diffeomorphism were also investigated, although in which the graviton was kept as transverse and traceless and thus the degrees of freedom are the same as in GR\footnote{It was argued in \cite{Khoury:2013oqa,Khoury:2014sea} that  general relativity is the unique spatially covariant effective field theory of the transverse and traceless graviton degrees of freedom.}.

Such spatially covariant gravity theories are most conveniently constructed using the Arnowitt-Deser-Misner (ADM) variables, i.e., the lapse $N$, shift $N_i$ and the spatial metric $h_{ij}$, which are adapted to such a space/time splitting nature.
In the study of EFT of inflation/dark energy, much attention was paid to polynomials of (perturbations of) lapse function $\delta g^{00} \equiv 2\delta N/N^3$ and the extrinsic curvature $\delta K^i_j$ with time-dependent parameters.
For the Ho\v{r}ava gravity, besides the linear combination $K_{ij}K^{ij}- \lambda K^2$, attention was mainly focused on higher order polynomials built of spatial curvature and its spatial derivatives such as ${}^{(3)}R^2$, ${}^{(3)}R_{ij}{}^{(3)}R^{ij}$, $h^{ij}\nabla_i {}^{(3)}R \nabla_j {}^{(3)}R$ etc, with constant parameters.
In the healthy extension of Ho\v{r}ava gravity \cite{Blas:2009qj}, terms such as $(a_ia^i)^n$ with $a_i =\partial_i \ln N$ were also introduced.
On the other hand, when rewriting the Horndeski theory in the unitary gauge \cite{Gleyzes:2013ooa}, couplings between the extrinsic and intrinsic curvatures (such as $K {}^{(3)}R$ and $K_{ij}{}^{(3)}R^{ij}$) and terms cubic in the extrinsic curvature naturally arise.
Very recently in \cite{Gleyzes:2014dya}, by deforming the Horndeski Lagrangian in the unitary gauge \cite{Gleyzes:2013ooa}, more general terms were introduced, which are special combinations of polynomials of the extrinsic curvature $K_{ij}$, with parameters being generalized as functions of $t$ and $N$.
Remarkably, when restoring the general covariance by introducing the St\"{u}ckelberg field $\phi$, the corresponding scalar tensor theory generally has higher order equations of motion, although by construction the number of degrees of freedom is kept up to three \cite{Gleyzes:2014dya,Gleyzes:2014qga}.
This feature significantly enlarges our understanding of scalar tensor theories beyond the Horndeski one.

Inspired by these studies, in \cite{Gao:2014soa} we proposed a framework for a wide class of spatially covariant gravity theories, which propagate at most three degrees of freedom.
The Lagrangians can be written in terms of polynomials of the extrinsic curvature $K_{ij}$, with coefficients being general functions of time $t$ and the spatial metric $h_{ij}$, the lapse function $N$ and the spatial Ricci tensor $R_{ij}$ as well as their spatial derivatives.
By construction, the framework has virtually included all the previous models.
Due to its generality, one may expect the existence of an even larger class of scalar tensor theories that have higher order equations of motion while still propagating three degrees of freedom.

Before applying our general framework \cite{Gao:2014soa} to cosmology such as modeling the inflation and dark energy, it is important the ensure that the theories themselves are consistent, especially, there are indeed at most three degrees of freedom are propagating.
Counting degrees of freedom can be well performed in the Hamiltonian constraint analysis. 
For the Lagrangian in \cite{Gleyzes:2014dya}, this has been done in \cite{Lin:2014jga} and \cite{Gleyzes:2014qga}, where terms up to quadratic order in the extrinsic curvature were considered.
Since the framework in \cite{Gao:2014soa} contains polynomials of the extrinsic curvature up to arbitrarily high orders, with coefficients including arbitrarily high order spatial derivatives of $N$ and $R_{ij}$, one might be concerned whether the theories have at most three degrees of freedom or not, or even how the Hamiltonian analysis could be performed.

This work is devoted to these issues.
We will keep full generalities of the framework in \cite{Gao:2014soa}, which we review in the next section.
The first difficulty in deriving the Hamiltonian of the theory is, due to the existence of arbitrarily high order powers of $K_{ij}$ in the Lagrangian, explicit solution for $K_{ij}$ in terms of the conjugate momenta $\pi^{ij}$ becomes impossible.
Nevertheless, in Sec. \ref{sec:Ham_cons} we formally give a series solution for $K_{ij}$, based on which the Hamiltonian can be expressed in terms of a polynomial of $\pi^{ij}$, with coefficients being functions of $t$, $h_{ij}$, $N$ and $R_{ij}$ as well as their arbitrary spatial derivatives.
In Sec. \ref{sec:PB_Phi_Ci}, we prove a general result on the Poisson bracket between the momentum constraint $\mathcal{C}_i$ and an arbitrary scalar density $\Phi$ of unit weight. The calculation of Poisson brackets among constraints and the counting number of degrees of freedom will be presented in Sec. \ref{sec:cons_alg}.

\smallskip{}

\emph{Notations}: Since we are dealing with spatial curvature terms, we omit their left superscript ${}^{(3)}$ for simplicity. $\nabla_i$ is the spatial covariant derivative, instead of the spatial components of a spacetime covariant derivative which we denote as $\nabla_{\mu}$. 
We sometimes use the following shorthands
	\begin{eqnarray*}
	\left\langle f,g\right\rangle  & \equiv & \int d^{3}x\, f\left(\vec{x}\right)g\left(\vec{x}\right),\\
	\left\langle f^{i},g_{i}\right\rangle  & \equiv & \int d^{3}x\, f^{i}\left(\vec{x}\right)g_{i}\left(\vec{x}\right),
	\end{eqnarray*}
etc, where $\left\langle \cdot,\cdot\right\rangle $ is symmetric with respect to its two arguments.

\section{The framework} \label{sec:framework}

Our purpose is to construct gravity theories which respect spatial diffeomorphism, and propagate no more than three degrees of freedom.
The basic ingredients in our construction are thus the lapse function $N$ defined in the ``3+1'' decomposition of the spacetime metric
	\begin{equation}
		\mathrm{d}s^2 = -N^2 \mathrm{d}t^2 + h_{ij} \left(\mathrm{d}x^i +N^i \mathrm{d}t\right) \left(\mathrm{d}x^j +N^j \mathrm{d}t\right), \label{ADM}
	\end{equation}
the intrinsic spatial curvature $R_{ij}$, as well as the extrinsic curvature $K_{ij}$ defined by
	\begin{equation}
		K_{ij} = \frac{1}{2 N}\left( \partial_t h_{ij} - \nabla_i N_j - \nabla_j N_i \right),
	\end{equation}
where $\nabla_i$ is the spatially covariant derivative compatible with the spatial metric $h_{ij}$.

Precisely, we consider a general class of Lagrangians of the following form \cite{Gao:2014soa}
	\begin{equation}
		\mathcal{L}= \sum_{n=1} \mathcal{K}_n \left[K\right]+ \mathcal{V}, \label{L_gen}
	\end{equation}
with 
	\begin{equation}
		\mathcal{K}_{n}\left[K\right]=\mathcal{G}_{(n)}^{i_{1}j_{1},\cdots,i_{n}j_{n}}K_{i_{1}j_{1}}\cdots K_{i_{n}j_{n}}, \label{L_kin}
	\end{equation}
where $\mathcal{G}_{(n)}$'s and $\mathcal{V}$ are general functions of
	\begin{equation}
		t,\quad h_{ij},\quad N,\quad R_{ij},\quad\nabla_{i}. \label{var_gen}
	\end{equation}
When writing (\ref{L_kin}), the symmetries of indices $(i_k j_k)$ of $\mathcal{G}_{(n)}$'s are understood.
The Lagrangian (\ref{L_gen}) describes gravity theories respecting the spatial diffeomorphism, which we may refer to as ``spatially covariant gravity''. 
Following the same strategy of \cite{Horava:2009uw}, it is convenient to view  $\mathcal{K}_n$ as the ``kinetic'' terms, since $K_{ij}$ involves first time derivative of the spatial metric $h_{ij}$, and $\mathcal{V}$ as the ``potential'' terms, respectively.

Please note in (\ref{L_gen}) we do not include the shift vector $N_i$ explicitly, although which is a vector under spatial diffeomorphism.
This can be understood as follows.
Geometrically, the Lagrangian (\ref{L_gen}) describes fluctuations of the foliation structure of spacetime, of which the spatial hypersurfaces are specified by the lapse function $N$. 
On the other hand, the shift vector $N_i$ by itself is not a genuine geometric quantity characterizing the foliation structure. 
Instead, it merely encodes the gauge freedom of spatial diffeomorphism, i.e., the freedom of choosing  coordinates on the spatial hypersurfaces. 
In fact, blindly including terms such as $N_i N^i$ would substantially change the constraint structure of the theory and inevitably introduce unwanted degrees of freedom.
In (\ref{L_gen}) the spatial Riemann tensor $R_{ijkl}$ does not appear, which is not an independent quantity since the spatial hypersurfaces are 3-dimensional.
Moreover, in this work we neglect spatial derivatives of $K_{ij}$ such as $\nabla_k K_{ij}$, $\nabla_k\nabla_l K_{ij}$ etc., although which may generally be allowed and interesting.

Coefficients of the kinetic terms $\mathcal{G}_{(n)}$'s and potential terms $\mathcal{V}$ have functional dependence on $N$, which implies the lapse function $N$ enters the theory nonlinearly (besides those implicitly through $K_{ij}$). 
This fact promotes $N$ from being a Lagrange multiplier associated with the gauge freedom of re-slicing the spacetime (as in GR) to an auxiliary variable.
In GR, the primary and secondary constraints associated with $N$ are both first class, which is the result of the space dependent time reparametrization invariance.
As we shall see in Sec. \ref{sec:cons_alg}, however, as long as $N$ enters the theory nonlinearly, as in our general construction (\ref{L_gen}), both the primary and secondary constraint associated with $N$ become second class, which is just the result of breaking full spacetime diffeomorphism to the spatial diffeomorphism. 
This fact crucially ensures the health of our construction, as well as the arising of a new degree of freedom comparing with GR.

By construction, the Lagrangian (\ref{L_gen}) is non-relativistic, since which is written in the ADM coordinates (\ref{ADM}), i.e., in a particular gauge (often referred to as the unitary gauge), and breaks general covariance explicitly. However, any non-relativistic theory can be thought as the  gauge fixed version of some relativistic theory, which can be got using the ``St\"{u}ckelberg trick''. 
In our case, the theory breaks general covariance by choosing a preferred foliation of spacelike hypersurfaces in the spacetime, thus the St\"{u}ckelberg field $\phi$ is introduced as the position of these hypersurfaces, with unit normal given by $n_{\mu}=\frac{-\nabla_{\mu}\phi}{\sqrt{2X}}$ with $X\equiv -(\nabla\phi)^2/2$. 
Then all the variables in (\ref{L_gen}) can be promoted to spacetime covariant versions, such as $N\rightarrow1/\sqrt{2X}$, and
	\begin{eqnarray}
	h_{ij} & \rightarrow & h_{\mu\nu}=g_{\mu\nu}+\frac{1}{2X}\nabla_{\mu}\phi\nabla_{\nu}\phi,\label{sr1}\\
	\nabla_{i}N & \rightarrow & h_{\mu}^{\nu}\nabla_{\nu}N\nonumber \\
	 & = & -\frac{1}{2\sqrt{2X}}\Big(\nabla_{\mu}\ln X+\frac{1}{2X}\nabla_{\mu}\phi\nabla^{\rho}\phi\nabla_{\rho}\ln X\Big),\quad \label{sr2}\\
	K_{ij} & \rightarrow & K_{\mu\nu}\nonumber \\
	 & = & -\frac{1}{\sqrt{2X}}\Big[\nabla_{\mu}\nabla_{\nu}\phi-\frac{1}{4X}\nabla_{\mu}\phi\nabla_{\nu}\phi\nabla_{\rho}\phi\nabla^{\rho}\ln X\nonumber \\
	 &  & \qquad \quad -\nabla_{(\mu}\phi\nabla_{\nu)}\ln X\Big],\label{sr3}
	\end{eqnarray}
etc. With these replacements (as well as the Gauss/Codazzi/Ricci equations), the Lagrangian (\ref{L_gen}) can be recast into a spacetime covariant theory describing the St\"{u}ckelberg field $\phi$ coupled to gravity.
Indeed, this is exactly the same technique in \cite{Germani:2009yt,Blas:2009yd,Blas:2010hb} (see also \cite{Blas:2009qj}) where Ho\v{r}ava gravity was reformulated in a fully covariant manner.

Conversely, starting from a generally covariant Lagrangian for a scalar field $\phi$ coupled to gravity
	\begin{equation}
		\mathcal{L} = \mathcal{L}(g_{\mu\nu},\phi, \nabla_\mu \phi, \nabla_\mu \nabla_\nu \phi, \cdots, {}^{(4)}R_{\mu\nu\rho\sigma},\cdots), \label{L_cov}
	\end{equation}
where ``$\cdots$'' denotes possible higher order covariant derivatives of $\phi$ and the curvature ${^{(4)}}R_{\mu\nu\rho\sigma}$, 
as long as the gradient of the scalar field is timelike, i.e. $g^{\mu\nu}\nabla_{\mu} \phi \nabla_{\nu}\phi <0$, we may choose the time coordinate as $t=\phi$, which (partially) fixes the gauge (unitary gauge).
In the unitary gauge, derivatives of the scalar field can be recast as, at the first order, $\nabla_{\mu}\phi=-n_{\mu}/N$ with $n_{\mu}=-N\delta_{\mu}^{0}$, and
at the second order, 
	\begin{equation}
	\nabla_{\mu}\nabla_{\nu}\phi=\frac{1}{N}\left(-n_{\mu}n_{\nu}\rho+2n_{(\mu}a_{\nu)}-K_{\mu\nu}\right),\label{d2phi}
	\end{equation}
where
	\begin{equation}
	\rho=\pounds_{\bm{n}}\ln N,\qquad a_{\mu}=e_{\mu}^{i}\partial_{i}\ln N,\qquad K_{\mu\nu}=e_{\mu}^{i} e_{\nu}^{j}K_{ij},\label{a_K}
	\end{equation}
with  $e_{\mu}^{i}=\delta_{\mu}^{0}N^{i}+\delta_{\mu}^{i}$, which can be expanded explicitly as
	\begin{eqnarray}
	\nabla_{\mu}\nabla_{\nu}\phi & = & -\frac{1}{N}\Big[\delta_{\mu}^{0}\delta_{\nu}^{0}\left(\partial_{t}N+N^{i}\partial_{i}N+N^{i}N^{j}K_{ij}\right)\nonumber \\
	 &  & +2\delta_{(\mu}^{0}\delta_{\nu)}^{i}\left(\partial_{i}N+N^{j}K_{ij}\right)+\delta_{\mu}^{i}\delta_{\nu}^{j}K_{ij}\Big].
	\end{eqnarray}
Similar procedures can be performed on higher orders. For example, taking a further derivative of (\ref{d2phi}) yields
	\begin{eqnarray}
	 &  & \nabla_{\mu}\nabla_{\nu}\nabla_{\rho}\phi\nonumber \\
	 & = & \frac{1}{N}\Big[n_{\mu}n_{\nu}n_{\rho}\left(-\rho^{2}+\pounds_{\bm{n}}\rho-2a^{\sigma}a_{\sigma}\right)\nonumber \\
	 &  & \quad-2n_{\mu}n_{(\nu}\big(\pounds_{\bm{n}}a_{\rho)}-2a_{\rho)}\rho-2K_{\rho)}^{\sigma}a_{\sigma}\big)\nonumber \\
	 &  & \quad+n_{\mu}\big(-2a_{\nu}a_{\rho}-\rho K_{\nu\rho}+\pounds_{\bm{n}}K_{\nu\rho}-2K_{\nu d}K_{\rho}^{d}\big)\nonumber \\
	 &  & \quad-\left(\pounds_{\bm{n}}a_{\mu}-2a_{\mu}\rho-2K_{\mu}^{\sigma}a_{\sigma}\right)n_{\nu}n_{\rho}\nonumber \\
	 &  & \quad+2\left(-a_{\mu}a_{(\nu}+D_{\mu}a_{(\nu}-\rho K_{\mu(\nu}-K_{\mu}^{\sigma}K_{\sigma(\nu}\right)n_{\rho)}\nonumber \\
	 &  & \quad+a_{\mu}K_{\nu\rho}+2K_{\mu(\nu}a_{\rho)}-D_{\mu}K_{\nu\rho}\Big],\label{d3phi}
	\end{eqnarray}
where $D_{\mu}$ is the spatially projected covariant derivative, e.g., $D_{\mu} a_{\nu} = e_{\mu}^{i} e_{\nu}^{j} \nabla_i a_j$ etc. (keep in mind that throughout this paper $\nabla_i$ always denotes covariant derivative associated with the spatial metric $h_{ij}$).
Using the above replacements, together with the Gauss/Codazzi/Ricci equations, the general Lagrangian (\ref{L_cov}) can be recast in terms of  $K_{ij}$, $R_{ij}$, $N$, $N^i$ as well as their temporal and spatial derivatives. 
For example, in \cite{Gleyzes:2013ooa}, the covariant Horndeski theory has been rewritten in the unitary gauge, where the corresponding Lagrangian falls into a subclass of (\ref{L_gen}).
For a general scalar-tensor theory (\ref{L_cov}), however, the corresponding Lagrangian in the unitary gauge may generally depend on the shift $N^i$  and  time derivatives of $K_{ij}$, $R_{ij}$ and $N$ (as well as their spatial derivatives), which signifies the existence of unwanted degrees of freedom, since the resulting equations of motion contain higher \emph{time} derivatives explicitly.
This is avoided in our construction (\ref{L_gen}), where the Lagrangian depends only on $K_{ij}$, $R_{ij}$, $N$ as well as their spatial derivatives.
The remarkable finding in \cite{Gleyzes:2014dya} is that, when reintroducing the St\"{u}ckelberg field $\phi$ as in (\ref{sr1})-(\ref{sr3}), the covariant Lagrangian for such kind of spatially covariant gravity theories corresponds to a class of scalar tensor theories (\ref{L_cov}), which have higher order equations of motion, although by construction the number of degrees of freedom is kept up to three.
This fact is thus beyond the scope of Horndeski theory, which requires the equations of motion to be second order.

As an explicit example of our general framework (\ref{L_gen})-(\ref{L_kin}), in \cite{Gao:2014soa} we proposed a ``cubic construction'', by imposing two further restrictions: 
1) there are no higher order derivatives in the Lagrangian when making the St\"{u}ckelberg replacement described above, i.e. we omit higher spatial derivatives of $R_{ij}$ and only keep the first derivative of the lapse $\nabla_i N$,  and 
2) the powers of second derivative operators do not exceed three.
This allows us to exhaust all the possible operators\footnote{Note here we slightly modified the Lagrangian in \cite{Gao:2014soa} by replacing $a_i=\nabla_i \ln N$ with $\nabla_i N$. In the former case, with constant parameters $a_n$, $b_n$ etc., the Lagrangian possesses an enhanced symmetry, i.e. space independent time reparametrization invariance $t\rightarrow \tilde{t}(t)$, as in the non-projectable version of Ho\v{r}ava gravity.}: 
for the ``kinetic terms''
	\begin{eqnarray}
	\mathcal{K}_{1} & = & \left(a_{0}+a_{1}R+a_{3}R^{2}+a_{4}R_{ij}R^{ij}+a_{5}\nabla_{i}N\nabla^{i}N\right)K\nonumber \\
	 &  & +[(a_{2}+a_{6}R)R_{j}^{i}+a_{7}R_{k}^{i}R_{j}^{k}+a_{8}\nabla^{i}N\nabla_{j}N]K_{i}^{j},\qquad \label{K1}\\
	\mathcal{K}_{2} & = & \left(b_{1}+b_{3}R\right)K^{2}+\left(b_{2}+b_{4}R\right)K_{ij}K^{ij}\nonumber \\
	 &  & +\left(b_{5}KK_{ij}+b_{6}K_{ik}K_{j}^{k}\right)R^{ij},\label{K2}\\
	\mathcal{K}_{3} & = & c_{1}K^{3}+c_{2}KK_{ij}K^{ij}+c_{3}K_{j}^{i}K_{k}^{j}K_{i}^{k},\label{K3}
	\end{eqnarray}
and for the ``potential terms''
	\begin{eqnarray}
	\mathcal{V} & = & d_{0}+d_{1}R+d_{2}R^{2}+d_{3}R_{ij}R^{ij}+d_{4}\nabla_{i}N\nabla^{i}N\nonumber \\
	 &  & +d_{5}R^{3}+d_{6}RR_{ij}R^{ij}+d_{7}R_{j}^{i}R_{k}^{j}R_{i}^{k}\nonumber \\
	 &  & +d_{8}R\nabla_{i}N\nabla^{i}N+d_{9}R_{ij}\nabla^{i}N\nabla^{j}N.\label{L_V}
	\end{eqnarray}
where $a_n,b_n,c_n,d_n$ are arbitrary functions of $t$ and $N$ (without derivatives).
This ``cubic construction'' has virtually included \emph{all} the previous models, while still possessing new interesting extensions.
The ``6-parameter'' Lagrangian presented in \cite{Gleyzes:2014dya} thus corresponds to
	\begin{align*}
	 & a_{0}=A_{3},\quad-2a_{1}=a_{2}=B_{5},\quad b_{1}=-b_{2}=A_{4},\\
	 & c_{1}=-\frac{1}{3}c_{2}=\frac{1}{2}c_{3}=A_{5},\quad d_{0}=A_{2},\quad d_{1}=B_{4},
	\end{align*}
with all other coefficients vanishing.

\section{Hamiltonian and constraints} \label{sec:Ham_cons}

The main purpose of this work is to show our theory (\ref{L_gen})-(\ref{L_kin}) is healthy in the sense that it does not propagate unwanted degree(s) of freedom other than the two tensor and one scalar modes.
Counting number of degrees of freedom can be well-performed in the Hamiltonian analysis.
In this section, we derive the Hamiltonian and the constraints of our theory.

The ten variables $\varphi^I \equiv \{N,N^i,h_{ij}\}$ as well as their conjugate momenta $\pi_I \equiv \{\pi_N, \pi_i, \pi^{ij}\}$ spanned a 20-dimensional phase space.
The explicit absence of time derivatives of the lapse $N$ and the shift $N_i$ in the Lagrangian implies the identical vanishing of their conjugate momenta $\pi_N$ and $\pi_i$, which yields 4 primary constraints
	\begin{equation}
		\pi_N = 0,\qquad \pi_i = 0, \label{cons_pri}
	\end{equation}
in the phase space.
The conjugate momenta to the spatial metric $h_{ij}$ are given by
	\begin{equation}
		\pi^{ij}\equiv \frac{\partial\big(N\sqrt{h}\mathcal{L}\big)}{\partial\dot{h}_{ij}} =\frac{\sqrt{h}}{2}\Pi^{ij}[K], \label{conj_mom}
	\end{equation}
where $h\equiv \det h_{ij}$, and for an arbitrary symmetric matrix $M_{ij}$, $\Pi^{ij}[M]$ is defined by
	\begin{equation}
		\Pi^{ij}[M]\equiv\mathcal{G}_{(1)}^{ij}+\sum_{n=1}(n+1)\mathcal{G}_{(n+1)}^{ij,k_{1}l_{1},\cdots,k_{n}l_{n}}M_{k_{1}l_{1}}\cdots M_{k_{n}l_{n}},  \label{Piij_def}
	\end{equation}
where $\mathcal{G}_{(n)}$'s are the same as in (\ref{L_kin}).
According to (\ref{conj_mom}), $\pi^{ij}$ itself is a spatial tensor density of unit weight (i.e. $\pi^{ij}/\sqrt{h}$ transforms as a spatial tensor).
For later convenience, we also define
	\begin{equation}
		\Pi^{ij,kl}[M]\equiv\frac{\partial\Pi^{ij}}{\partial M_{kl}},\quad\Pi^{ij,k_{1}l_{1},k_{2}l_{2}}[M]\equiv\frac{\partial^{2}\Pi^{ij}}{\partial M_{k_{1}l_{1}}\partial M_{k_{2}l_{2}}},
	\end{equation}
etc. for short.

In the case of GR, 
	\[
		\Pi^{ij}[K]=M_{\mathrm{pl}}^{2}\left(K^{ij}-h^{ij}K\right),
	\]
which is linear in $K_{ij}$ and thus the inversion of $K_{ij}$ (equivalently, $\dot{h}_{ij}$) in terms of $\pi^{ij}$ can be done easily.
In the study of Hamiltonian analysis in \cite{Lin:2014jga,Gleyzes:2014qga}, only terms up to quadratic power of $K_{ij}$ in the Lagrangian are considered, which also implies $\Pi^{ij}[K]$ is linear in $K_{ij}$, and thus the inversion can also be easily made.
Generally (\ref{conj_mom}) is a nonlinear algebraic equation for $K_{ij}$, from which solving $K_{ij}$ in terms of $\pi^{ij}$ may be involved.
Nevertheless, we make a general ansatz for the solution of (\ref{conj_mom}), in terms of series of $\pi^{ij}$:
	\begin{equation}
		K_{ij}=\Gamma_{ij}^{(1)}+\frac{1}{\sqrt{h}}\Gamma_{ij,kl}^{(2)}\pi^{kl}+\frac{1}{h}\Gamma_{ij,k_{1}l_{1},k_{2}l_{2}}^{(3)}\pi^{k_{1}l_{1}}\pi^{k_{2}l_{2}}+\cdots, \label{K_ansatz}
	\end{equation}
where the coefficients $\Gamma^{(n)}$'s are also generally functions of $t$, $h_{ij}$, $N$ and $R_{ij}$ as well as their spatial derivatives.
In writing (\ref{K_ansatz}), we have deliberately separated powers of $h\equiv \det h_{ij}$ in the denominators such that $\Gamma^{(n)}$'s are spatially covariant tensors.
The coefficients $\Gamma^{(n)}$'s can be solved perturbatively by plugging (\ref{K_ansatz}) into (\ref{conj_mom}), which yields
	\begin{eqnarray}
	2\frac{\pi^{ij}}{\sqrt{h}} & = & \Pi^{ij}[\Gamma^{(1)}]+\frac{\pi^{i_{1}j_{1}}}{\sqrt{h}}\Pi^{ij,kl}[\Gamma^{(1)}]\Gamma_{kl,i_{1}j_{1}}^{(2)}\nonumber \\
	 &  & +\frac{\pi^{i_{1}j_{1}}\pi^{i_{2}j_{2}}}{h}\Big(\Pi^{ij,kl}[\Gamma^{(1)}]\,\Gamma_{kl,i_{1}j_{1},i_{2}j_{2}}^{(3)}\nonumber \\
	 &  & +\frac{1}{2}\Pi^{ij,k_{1}l_{1},k_{2}l_{2}}[\Gamma^{(1)}]\,\Gamma_{k_{1}l_{1},i_{1}j_{1}}^{(2)}\Gamma_{k_{2}l_{2},i_{2}j_{2}}^{(2)}\Big)\nonumber \\
	 &  & +\frac{\pi^{i_{1}j_{1}}\pi^{i_{2}j_{2}}\pi^{i_{3}j_{3}}}{h^{3/2}}\Big(\Pi^{ij,kl}[\Gamma^{(1)}]\,\Gamma_{kl,i_{1}j_{1},i_{2}j_{2},i_{3}j_{3}}^{(4)}\nonumber \\
	 &  & +\Pi^{ij,k_{1}l_{1},k_{2}l_{2}}[\Gamma^{(1)}]\,\Gamma_{k_{1}l_{1},i_{1}j_{1}}^{(2)}\Gamma_{k_{2}l_{2},i_{2}j_{2},i_{3}j_{3}}^{(3)}\nonumber \\
	 &  & +\frac{1}{6}\Pi^{ij,k_{1}l_{1},k_{2}l_{2},k_{3}l_{3}}[\Gamma^{(1)}]\nonumber \\
	 &  & \times\Gamma_{k_{1}l_{1},i_{1}j_{1}}^{(2)}\Gamma_{k_{2}l_{2},i_{2}j_{2}}^{(2)}\Gamma_{k_{3}l_{3},i_{3}j_{3}}^{(2)}\Big)+\mathcal{O}(\pi^{4}/h^{2}).\quad\label{Gamma_eq}
	\end{eqnarray}
Comparing both sides of (\ref{Gamma_eq}), $\Gamma^{(1)}_{ij}$ is solved by
	\begin{equation}
		\Pi^{ij}[\Gamma^{(1)}] = 0,
	\end{equation}
where recall that $\Pi^{ij}[M]$ is defined in (\ref{Piij_def}).
$\Gamma_{ij,kl}^{(2)}$ can be  determined by
	\begin{equation}
		\Gamma_{ij,kl}^{(2)}=2\Pi_{ij,kl}^{-1}[\Gamma^{(1)}],
	\end{equation}
where $\Pi_{ij,kl}^{-1}[\Gamma^{(1)}]$ is the inverse of $\Pi^{ij,kl}[\Gamma^{(1)}]$ satisfying\footnote{The existence of the inverse $\Pi_{ij,kl}^{-1}[\Gamma^{(1)}]$ is crucial. Otherwise the kinetic term for $h_{ij}$ is degenerate and there are additional primary constraints besides the ones in (\ref{cons_pri}). The later is similar to the Ho\v{r}ava gravity with $\lambda=1/3$, where it was argued that (e.g. \cite{Bellorin:2013zbp}) the theory propagates two physical degrees of freedom due to the additional primary constraint $h_{ij}\pi^{ij} =0$ as well as its associated secondary constraint.}
	\begin{equation}
			\Pi^{ij,k'l'}[\Gamma^{(1)}]\Pi_{k'l',kl}^{-1}[\Gamma^{(1)}]=\mathbf{I}_{kl}^{ij},
		\end{equation}
with $\mathbf{I}_{kl}^{ij}$ the identity in the space of symmetric matrices $\mathbf{I}_{kl}^{ij}\equiv\delta_{(k}^{i}\delta_{l)}^{j}$.
The vanishing of terms nonlinear in $\pi^{ij}$ on the right-hand-side of (\ref{Gamma_eq}) thus yields a hierarchy of equations, from which we may solve
	\begin{eqnarray}
	\Gamma_{i_{1}j_{1},i_{2}j_{2},i_{3}j_{3}}^{(3)} & = & -2\Pi^{k_{1}l_{1},k_{2}l_{2},k_{3}l_{3}}[\Gamma^{(1)}]\,\Pi_{k_{1}l_{1},i_{1}j_{1}}^{-1}[\Gamma^{(1)}] \nonumber\\
	 &  & \times\Pi_{k_{2}l_{2},i_{2}j_{2}}^{-1}[\Gamma^{(1)}]\Pi_{k_{3}l_{3},i_{3}j_{3}}^{-1}[\Gamma^{(1)}],
	\end{eqnarray}
etc.
Following this procedure, one may solve  all the coefficients $\Gamma^{(n)}$'s in (\ref{K_ansatz}) in terms of $\mathcal{G}_{(n)}$'s and $\mathcal{V}$ in (\ref{L_gen}), at least in principle.
From now on, we use (\ref{K_ansatz}) as our starting point, by assuming all the coefficients $\Gamma^{(n)}$'s have been determined as functions of $t$, $h_{ij}$, $N$ and $R_{ij}$ as well as their spatially covariant derivatives.
As we shall see, fortunately, the concrete expressions for the solutions are not necessary for our purpose.

Due to the presence of the 4 primary constraints, the canonical Hamiltonian can be extended arbitrarily off the 16-dimensional hypersurface in phase space specified by the primary constraints (\ref{cons_pri}), which yields the so-called ``total Hamiltonian'':
	\begin{eqnarray}
	H_{\mathrm{T}} & \equiv & \int d^{3}x\left(\pi^{ij}\dot{h}_{ij}-N\sqrt{h}\mathcal{L}+\lambda^{N}\pi_{N}+\lambda^{i}\pi_{i}\right)\nonumber \\
	 & \simeq & \int d^{3}x\left(N\tilde{\mathcal{C}}+N_{i}\mathcal{C}^{i}+\lambda^{N}\pi_{N}+\lambda^{i}\pi_{i}\right),\label{H_total}
	\end{eqnarray}
where where $\lambda^{N}$ and $\lambda^{i}$ are Lagrange multipliers associated
with the primary constraints, and
	\begin{eqnarray}
	\tilde{\mathcal{C}} & = & 2\pi^{ij}K_{ij}-\sqrt{h}\mathcal{L},\label{C_tilde}\\
	\mathcal{C}^{i} & = & -2\sqrt{h}\nabla_{j}\left(\frac{\pi^{ij}}{\sqrt{h}}\right).\label{Ci}
	\end{eqnarray}
At this point, it is important to notice that due to the solution (\ref{K_ansatz}), $\tilde{\mathcal{C}}$ is generally a polynomial of $\pi^{ij}$, with coefficients being functions of $t$, $h_{ij}$, $N$ and $R_{ij}$ as well as their spatial derivatives.
Note $\mathcal{C}^i$ is irrelevant to any specific structure of our theory and is exactly the same as that of GR, which is the result of spatial invariance of the theory. 
On the other hand, $\tilde{\mathcal{C}}$ is subject to the concrete form of the Lagrangian (\ref{L_gen}) and thus varies from model to model.
Note both $\mathcal{C}$ and $\tilde{\mathcal{C}}$ have nothing to do with the shift vector $N^i$.

For arbitrary functions $A$ and $B$ of canonical variables $(\varphi^I,\pi_I)$, the Poisson bracket is defined by
	\begin{equation}
		\left\{ A,B\right\} _{\mathrm{P}}\equiv\sum_{I}\int d^{3}x\bigg(\frac{\delta A}{\delta\varphi^{I}(\vec{x})}\frac{\delta B}{\delta\pi_{I}(\vec{x})}-\frac{\delta A}{\delta\pi_{I}(\vec{x})}\frac{\delta B}{\delta\varphi^{I}(\vec{x})}\bigg). \label{PB_def}
	\end{equation}
The time evolution of any function $F=F\left(t,\varphi^{I},\pi_{I}\right)$ defined on the phase space is thus given by
	\begin{equation}
		\frac{\mathrm{d}F}{\mathrm{d}t} \approx \frac{\partial F}{\partial t}+\left\{ F,H_{\mathrm{T}}\right\} _{\mathrm{P}} , \label{t_evo}
	\end{equation}
where the total Hamiltonian $H_{\mathrm{T}}$ is given in (\ref{H_total}),
and ``$\approx$'' denotes the ``weak equality'' that holds when the primary constraints (\ref{cons_pri}) are satisfied.

Constraints must be preserved in time.
By evaluating the time evolution of the primary constraints $\pi_N = 0$ and $\pi_i= 0$, we get
	\begin{eqnarray}
	\frac{\mathrm{d}}{\mathrm{d}t}\pi_{N} & \approx & \left\{ \pi_{N},H_{\mathrm{T}}\right\} _{\mathrm{P}}=-\mathcal{C},\label{piN_dt}\\
	\frac{\mathrm{d}}{\mathrm{d}t}\pi_{i} & \approx & \left\{ \pi_{i},H_{\mathrm{T}}\right\} _{\mathrm{P}}=-\mathcal{C}_{i},\label{pii_dt}
	\end{eqnarray}
with\footnote{In the case of $\tilde{\mathcal{C}}$ including $N$ only (instead of derivatives of $N$), (\ref{C_def}) reduces to $\mathcal{C}=\tilde{\mathcal{C}}+N\frac{\partial\tilde{\mathcal{C}}}{\partial N}\equiv\frac{\partial\left(N\tilde{\mathcal{C}}\right)}{\partial N}$, which is the case considered in \cite{Lin:2014jga,Gleyzes:2014qga}.}
	\begin{align}
	\mathcal{C}  \equiv & \frac{\delta}{\delta N}\int d^{3}x\left(N\tilde{\mathcal{C}}\right)\nonumber \\
	  = & \tilde{\mathcal{C}}+\sqrt{h}\sum_{n=0}(-1)^{n}\nabla_{i_{n}}\cdots\nabla_{i_{1}}\bigg(\frac{N}{\sqrt{h}}\frac{\partial\tilde{\mathcal{C}}}{\partial(\nabla_{i_{1}}\cdots\nabla_{i_{n}}N)}\bigg),\label{C_def}
	\end{align}
where $\tilde{\mathcal{C}}$ and  $\mathcal{C}_i$ are given in (\ref{C_tilde})-(\ref{Ci}).
In (\ref{C_def}), the case $n=0$ in the summation simply denotes the term $N \frac{\partial \tilde{\mathcal{C}}}{\partial N}$ without spatial derivatives.
If $\tilde{\mathcal{C}}$ has no functional dependence on $N$, (\ref{C_def}) implies $\mathcal{C}= \tilde{\mathcal{C}}$, which is just the case of GR.
The vanishing of (\ref{piN_dt})-(\ref{pii_dt}) corresponds to the so-called secondary constraints.
Together with the primary constraints, we are thus left with totally 8 constraints
	\begin{equation}
		\pi_N \approx 0,\qquad \pi_i \approx 0, \qquad  \mathcal{C} \approx 0, \qquad \mathcal{C}_i \approx 0. \label{cons_all}
	\end{equation}
Now the total Hamiltonian (\ref{H_total}) should also be supplemented by a term $\lambda_{\mathcal{C}}\mathcal{C}$ in the integrand, which yields the so-called ``extended Hamiltonian'':
	\begin{equation}
		H_{\mathrm{E}}\equiv\int d^{3}x\left(N\tilde{\mathcal{C}}+N_{i}\mathcal{C}^{i}+\lambda^{N}\pi_{N}+\lambda^{i}\pi_{i}+\lambda_{\mathcal{C}}\mathcal{C}\right). \label{H_ex}
	\end{equation}
In (\ref{cons_all}) and in what follows, ``$\approx$'' denotes the ``weak equality'' that holds on the constraint surface specified by (\ref{cons_all}).
As we will discuss later in Sec. \ref{sec:cons_alg}, although $\mathrm{d}\mathcal{C}_i/\mathrm{d} t \not\approx 0$, the linear combination $\mathcal{C}_i+ \pi_N\nabla_i N$ is automatically preserved on the constraint surface. While although $\mathrm{d}\mathcal{C}/\mathrm{d}t$ does not vanish  on the constraint surface, requiring $\mathrm{d}\mathcal{C}/\mathrm{d}t \approx 0$ (together with $\mathrm{d}\pi_N/\mathrm{d}t \approx 0$) merely fixes the Lagrange multipliers $\lambda^N$ and $\lambda_{\mathcal{C}}$ instead of generating new constraints. 
Thus the 8 constraints in (\ref{cons_all}) are  all the constraints in our theory.

\section{Poisson bracket $\left\{\Phi ,\mathcal{C}_{i}\right\} _{\mathrm{P}}$} \label{sec:PB_Phi_Ci}

Before evaluating the Poisson brackets among various constraints in (\ref{cons_all}) and counting the number of degrees of freedom, in this section we concentrate on the Poisson bracket between $\mathcal{C}_i$ and a general scalar density $\Phi$ of unit weight (i.e. $\Phi/\sqrt{h}$ is a scalar under spatial diffeomorphism), which encodes the action of $\mathcal{C}_i$ on $\Phi$. 
Precisely, we will prove that for a general scalar density $\Phi$ of unit weight defined on the phase space, which can be written in terms of a general polynomial of $\pi^{ij}$ with coefficients being functions of $t$, $h_{ij}$, $N$ and $R_{ij}$ as well as their spatial covariant derivatives such as $\nabla_i N$, $\nabla_i \nabla_j N$, $\nabla_k R_{ij}$, $\nabla_k\nabla_l R_{ij}$ etc, the following equality holds (up to surface terms): 
	\begin{equation}
		\{\Phi(\vec{x}'),\mathcal{C}_{i}(\vec{x})\}_{\mathrm{P}}=-\Phi(\vec{x})\nabla_{i}\delta^{3}(\vec{x}-\vec{x}')-\frac{\delta\Phi(\vec{x}')}{\delta N(\vec{x})}\nabla_{i}N(\vec{x}), \label{PB_Phi_Ci}
	\end{equation}
where the spatial functional derivative of $\Phi\left(\vec{x}'\right)$ with respect to $N\left(\vec{x}\right)$ is given explicitly by
	\begin{eqnarray}
	 &  & \frac{\delta\Phi(\vec{x}')}{\delta N(\vec{x})}\nonumber \\
	 & = & \delta^{3}\left(\vec{x}-\vec{x}'\right)\frac{\partial\Phi}{\partial N}+\sqrt{h}\sum_{n=1}\left(-1\right)^{n}\nonumber \\
	 &  & \times\nabla_{i_{n}}\cdots\nabla_{i_{1}}\bigg(\frac{\delta^{3}(\vec{x}-\vec{x}')}{\sqrt{h}}\frac{\partial\Phi}{\partial(\nabla_{i_{1}}\cdots\nabla_{i_{n}}N)}\bigg).\label{fd_Phi_N}
	\end{eqnarray}
(\ref{PB_Phi_Ci}) is one of the main results in this work.

In the following we show the derivation of (\ref{PB_Phi_Ci}) explicitly, which is lengthy, technical, but never straightforward. Although we prefer to put it in the main text in order to show the key steps, reads who are not interested in the mathematical details may skip to the next section.

First note that both sides of (\ref{PB_Phi_Ci}) are linear in $\Phi$. Since we assume $\Phi$ can be written in terms of a polynomial of $\pi^{ij}$, we are able to focus on a single monomial of the prototype
		\begin{equation}
			\Phi=\frac{1}{h^{\frac{n-1}{2}}}\Gamma_{i_{1}j_{1},\cdots,i_{n}j_{n}}\pi^{i_{1}j_{1}}\cdots\pi^{i_{n}j_{n}}, \label{Phi_mono}
		\end{equation}
where $h\equiv \det h_{ij}$, $\Gamma_{i_{1}j_{1},\cdots,i_{n}j_{n}}$ is a tensorial function of $t$, $h_{ij}$, $N$ and $R_{ij}$ as well as their spatially covariant derivatives. 
Here the factor $1/h^{\frac{n-1}{2}}$ is present since $\Phi$ is a scalar density of unit weight.
As long as (\ref{Phi_mono}) is proved to satisfy (\ref{PB_Phi_Ci}), (\ref{PB_Phi_Ci}) holds for a general polynomial.

In order to simplify the calculations of functional derivatives in the Poisson bracket, it is more convenient to evaluate the Poisson bracket between $\left\langle f,\Phi\right\rangle $ and $\left\langle g^{i},\mathcal{C}_{i}\right\rangle $ 
where $f$ and $g^i$ are test functions of spatial coordinates only and satisfy $f,g^i \xrightarrow{|\vec{x}|\rightarrow \infty} 0$, which
allows us to eliminate all boundary terms when performing integrations by parts.
By definition
	\begin{eqnarray}
	 &  & \left\{ \left\langle f,\Phi\right\rangle ,\left\langle g^{i},\mathcal{C}_{i}\right\rangle \right\} _{\mathrm{P}}\nonumber \\
	 & = & \int d^{3}x\left(\frac{\delta\left\langle f,\Phi\right\rangle }{\delta h_{ij}\left(\vec{x}\right)}\frac{\delta\left\langle g^{k},\mathcal{C}_{k}\right\rangle }{\delta\pi^{ij}\left(\vec{x}\right)}-\frac{\delta\left\langle f,\Phi\right\rangle }{\delta\pi^{ij}\left(\vec{x}\right)}\frac{\delta\left\langle g^{k},\mathcal{C}_{k}\right\rangle }{\delta h_{ij}\left(\vec{x}\right)}\right)\nonumber \\
	 & = & \int d^{3}x\bigg[\frac{\delta\left\langle f,\Phi\right\rangle }{\delta h_{ij}}2\nabla_{(i}g_{j)}\nonumber \\
	 &  & -\frac{\delta\left\langle f,\Phi\right\rangle }{\delta\pi^{ij}}\left(2\nabla_{k}g^{(i}\pi^{j)k}-\sqrt{h}\nabla_{k}\left(g^{k}\frac{\pi^{ij}}{\sqrt{h}}\right)\right)\bigg],\label{PB_Phi_Ci_ori}
	\end{eqnarray}
where the functional derivatives of $\left\langle g^{k},\mathcal{C}_{k}\right\rangle $ with respect to $h_{ij}$ and $\pi^{ij}$ are the same as in GR (see Appendix \ref{sec:fd_Ci} for a brief derivation). 
Varying (\ref{Phi_mono}) with respect to $\pi^{ij}$ simply yields
	\begin{equation}
	\frac{\delta\left\langle f,\Phi\right\rangle }{\delta\pi^{ij}}=nf\frac{1}{h^{\frac{n-1}{2}}}\Gamma_{ij,i_{2}j_{2},\cdots,i_{n}j_{n}}\pi^{i_{2}j_{2}}\cdots\pi^{i_{n}j_{n}} \equiv f\frac{\partial\Phi}{\partial\pi^{ij}}.\label{fd_Phi_pi}
	\end{equation}
The functional derivative of $\left\langle f,\Phi\right\rangle $ with respect to $h_{ij}$ is much involved, though which is evaluated in Sec.\ref{sec:fd_Phi} and is given by
	\begin{equation}
		\frac{\delta\left\langle f,\Phi\right\rangle }{\delta h_{ij}}=-\frac{n-1}{2}f\,\Phi\, h^{ij}+\sqrt{h}\left(\Delta_{1}^{ij}+\Delta_{2}^{ij}\right), \label{fd_Phi_hij}
	\end{equation}
where $\Delta_{1}^{ij}$ and $\Delta_{2}^{ij}$ are defined in (\ref{Deltaij_1}) and (\ref{Deltaij_2}), respectively.
Plugging (\ref{fd_Phi_pi}) and (\ref{fd_Phi_hij}) into (\ref{PB_Phi_Ci_ori}), we have
	\begin{eqnarray}
	 &  & \left\{ \left\langle f,\Phi\right\rangle ,\left\langle g^{i},\mathcal{C}_{i}\right\rangle \right\} _{\mathrm{P}}\nonumber \\
	 & = & \int d^{3}x\Big[2\sqrt{h}\big(\Delta_{1}^{ij}+\Delta_{2}^{ij}\big)\nabla_{i}g_{j}-\left(n-1\right)f\Phi\nabla_{i}g^{i}\nonumber \\
	 &  & -2f\frac{\partial\Phi}{\partial\pi^{ij}}\nabla_{k}g^{i}\pi^{jk}+f\frac{\partial\Phi}{\partial\pi^{ij}}\sqrt{h}\nabla_{k}g^{k}\frac{\pi^{ij}}{\sqrt{h}}\nonumber \\
	 &  & +\frac{nf}{h^{\frac{n-1}{2}}}\Gamma_{ij,i_{2}j_{2},\cdots,i_{n}j_{n}}\pi^{i_{2}j_{2}}\cdots\pi^{i_{n}j_{n}}\sqrt{h}g^{k}\nabla_{k}\Big(\frac{\pi^{ij}}{\sqrt{h}}\Big)\Big]\nonumber \\
	 & = & \int d^{3}x\Big[2\sqrt{h}\big(\Delta_{1}^{ij}+\Delta_{2}^{ij}\big)\nabla_{i}g_{j}\nonumber \\
	 &  & +f\Phi\nabla_{i}g^{i}-2f\pi^{jk}\frac{\partial\Phi}{\partial\pi^{ij}}\nabla_{k}g^{i}\nonumber \\
	 &  & +f\Gamma_{ij,i_{2}j_{2},\cdots,i_{n}j_{n}}\sqrt{h}g^{k}\nabla_{k}\Big(\frac{1}{h^{\frac{n}{2}}}\pi^{ij}\pi^{i_{2}j_{2}}\cdots\pi^{i_{n}j_{n}}\Big)\Big],\qquad \label{step_1}
	\end{eqnarray}
where in arriving at the second equality we used
	\[
		\frac{\partial\Phi}{\partial\pi^{ij}}\pi^{ij}=n\frac{1}{h^{\frac{n-1}{2}}}\Gamma_{ij,i_{2}j_{2},\cdots,i_{n}j_{n}}\pi^{i_{2}j_{2}}\cdots\pi^{i_{n}j_{n}}\pi^{ij}\equiv n\Phi,
	\]
and
	\begin{eqnarray*}
	 &  & \Gamma_{ij,i_{2}j_{2},\cdots,i_{n}j_{n}}\frac{1}{h^{\frac{n-1}{2}}}\pi^{i_{2}j_{2}}\cdots\pi^{i_{n}j_{n}}\nabla_{k}\left(\frac{\pi^{ij}}{\sqrt{h}}\right)\\
	 & = & \frac{1}{n}\Gamma_{i_{1}j_{1},\cdots,i_{n}j_{n}}\nabla_{k}\left(\frac{1}{h^{\frac{n}{2}}}\pi^{i_{1}j_{1}}\cdots\pi^{i_{n}j_{n}}\right).
	\end{eqnarray*}
Integrating by parts the second line of (\ref{step_1}) and using the definition for $\Phi$ again yield
	\begin{eqnarray}
	 &  & \left\{ \left\langle f,\Phi\right\rangle ,\left\langle g^{i},\mathcal{C}_{i}\right\rangle \right\} _{\mathrm{P}}\nonumber \\
	 & \simeq & \int d^{3}x\Big[2\sqrt{h}(\Delta_{1}^{ij}+\Delta_{2}^{ij})\nabla_{i}g_{j}-2f\pi^{jk}\frac{\partial\Phi}{\partial\pi^{ij}}\nabla_{k}g^{i}\nonumber \\
	 &  & -g^{k}\nabla_{k}f\Phi-\frac{fg^{k}}{h^{\frac{n-1}{2}}}\nabla_{k}\Gamma_{i_{1}j_{1},\cdots,i_{n}j_{n}}\pi^{i_{1}j_{1}}\cdots\pi^{i_{n}j_{n}}\Big].\qquad \label{step_2}
	\end{eqnarray}
For the last term in (\ref{step_2}), since $\Gamma_{i_{1}j_{1},\cdots,i_{n}j_{n}}$ is a function of $N$ and $R_{ij}$ and their spatial derivatives, we have
	\begin{eqnarray}
	 &  & \frac{fg^{k}}{h^{\frac{n-1}{2}}}\nabla_{k}\Gamma_{i_{1}j_{1},i_{2}j_{2},\cdots,i_{n}j_{n}}\pi^{i_{1}j_{1}}\cdots\pi^{i_{n}j_{n}}\nonumber \\
	 & = & fg^{k}\bigg(\sum_{n'=0}^{2}\nabla_{k}\nabla_{m_{1}}\cdots\nabla_{m_{n'}}R_{ij}\frac{\partial\Phi}{\partial(\nabla_{m_{1}}\cdots\nabla_{m_{n'}}R_{ij})}\nonumber \\
	 &  & +\sum_{n''=0}^{4}\nabla_{k}\nabla_{l_{1}}\cdots\nabla_{l_{n''}}N\frac{\partial\Phi}{\partial(\nabla_{l_{1}}\cdots\nabla_{l_{n''}}N)}\bigg).\label{nabla_Gamma}
	\end{eqnarray}
where the case $n'=0$ and $n''=0$ simply denote $R_{ij}$ and $N$ without derivatives.
Note in (\ref{nabla_Gamma}) we have truncated the derivatives of $R_{ij}$ up to the second order, and of $N$ up to the fourth order.
Plugging (\ref{nabla_Gamma}) into (\ref{step_2}) and after some manipulations, we have
	\begin{equation}
	\left\{ \left\langle f,\Phi\right\rangle ,\left\langle g^{i},\mathcal{C}_{i}\right\rangle \right\} _{\mathrm{P}}=-\int d^{3}x\, g^{k}\nabla_{k}f\,\Phi+\mathcal{I}_{1}+\mathcal{I}_{2},\label{PB_Phi_Ci_int}
	\end{equation}
where
	\begin{align}
	\mathcal{I}_{1}  \equiv & \int d^{3}x\bigg(2\sqrt{h}\Delta_{1}^{ij}\nabla_{i}g_{j}-2f\pi^{jk}\frac{\partial\Phi}{\partial\pi^{ij}}\nabla_{k}g^{i}\nonumber \\
	   & -fg^{k}\sum_{n=0}^{2}\nabla_{k}\nabla_{m_{1}}\cdots\nabla_{m_{n}}R_{ij}  \frac{\partial\Phi}{\partial(\nabla_{m_{1}}\cdots\nabla_{m_{n}}R_{ij})}\bigg),\label{I1_def}
	\end{align}
and
	\begin{eqnarray}
	\mathcal{I}_{2} & = & \int d^{3}x\bigg(2\sqrt{h}\Delta_{2}^{ij}\nabla_{i}g_{j}\nonumber \\
	 &  & -fg^{k}\sum_{n=0}^{4}\nabla_{k}\nabla_{i_{1}}\cdots\nabla_{i_{n}}N\frac{\partial\Phi}{\partial(\nabla_{i_{1}}\cdots\nabla_{i_{n}}N)}\bigg).\qquad \label{I2_def}
	\end{eqnarray}
	
Now our task is to calculate $\mathcal{I}_1$ and $\mathcal{I}_2$.
To this end, first plugging the explicit expression for $\Delta_{1}^{ij}$ (\ref{Deltaij_1}) into the first term of (\ref{I1_def}), and performing integrations by part to move all the covariant derivatives onto $\nabla_i g_j$ yield
	\begin{eqnarray}
	 &  & \int d^{3}x\,2\sqrt{h}\Delta_{1}^{ij}\nabla_{i}g_{j}\nonumber \\
	 & \simeq & \int d^{3}x\bigg[2f\left(\frac{\partial\Phi}{\partial h_{ij}}-h^{ik}h^{jl}\frac{\partial\Phi}{\partial h^{kl}}\right)\nabla_{i}g_{j}\nonumber \\
	 &  & +2f\frac{\partial\Phi}{\partial R_{kl}}\mathcal{A}_{kl}^{ijl_{1}l_{2}}\nabla_{l_{1}}\nabla_{l_{2}}\nabla_{i}g_{j}\nonumber \\
	 &  & +2f\frac{\partial\Phi}{\partial(\nabla_{m}R_{kl})}\nonumber \\
	 &  & \times\Big(\mathcal{A}_{kl}^{ijl_{1}l_{2}}\nabla_{m}\nabla_{l_{1}}\nabla_{l_{2}}\nabla_{i}g_{j}-\mathcal{B}_{mkl}^{ijl'}\nabla_{l'}\nabla_{i}g_{j}\Big)\nonumber \\
	 &  & +2f\frac{\partial\Phi}{\partial\left(\nabla_{m}\nabla_{n}R_{kl}\right)}\Big(\mathcal{A}_{kl}^{ijl_{1}l_{2}}\nabla_{m}\nabla_{n}\nabla_{l_{1}}\nabla_{l_{2}}\nabla_{i}g_{j}\nonumber \\
	 &  & -\mathcal{B}_{nkl}^{ijl'}\nabla_{m}\nabla_{l'}\nabla_{i}g_{j}-\mathcal{C}_{mnkl}^{ijl'}\nabla_{l'}\nabla_{i}g_{j}\Big)\bigg],\label{int_Delta_1}
	\end{eqnarray}
where tensors $\mathcal{A}$, $\mathcal{B}$ and $\mathcal{C}$ are defined in (\ref{Acal})-(\ref{Ccal}).
Then straightforward although tedious calculations show that
	\begin{equation}
	2\mathcal{A}_{kl}^{ijl_{1}l_{2}}\nabla_{l_{1}}\nabla_{l_{2}}\nabla_{i}g_{j}=g^{i}\nabla_{i}R_{kl}+2h_{(k}^{i}R_{l)}^{j}\nabla_{i}g_{j},\label{ctr_1}
	\end{equation}
and
	\begin{eqnarray}
	 &  & 2\left(\mathcal{A}_{kl}^{ijl_{1}l_{2}}\nabla_{m}\nabla_{l_{1}}\nabla_{l_{2}}\nabla_{i}g_{j}-\mathcal{B}_{mkl}^{ijl'}\nabla_{l'}\nabla_{i}g_{j}\right)\nonumber \\
	 & = & g^{i}\nabla_{i}\nabla_{m}R_{kl}\nonumber \\
	 &  & +\left(h_{l}^{i}\nabla_{m}R_{k}^{j}+h_{k}^{i}\nabla_{m}R_{l}^{j}+h_{m}^{i}\nabla^{j}R_{kl}\right)\nabla_{i}g_{j}.\label{ctr_2}
	\end{eqnarray}	
and
	\begin{eqnarray}
	 &  & 2\Big(\mathcal{A}_{kl}^{ijl_{1}l_{2}}\nabla_{m}\nabla_{n}\nabla_{l_{1}}\nabla_{l_{2}}\nabla_{i}g_{j}\nonumber \\
	 &  & -\mathcal{B}_{nkl}^{ijl'}\nabla_{m}\nabla_{l'}\nabla_{i}g_{j}-\mathcal{C}_{mnkl}^{ijl'}\nabla_{l'}\nabla_{i}g_{j}\Big)\nonumber \\
	 & = & g^{i}\nabla_{i}\nabla_{m}\nabla_{n}R_{kl}+\Big(h_{n}^{i}\nabla_{m}\nabla^{j}R_{kl}+h_{m}^{i}\nabla^{j}\nabla_{n}R_{kl}\nonumber \\
	 &  & +h_{l}^{i}\nabla_{m}\nabla_{n}R_{k}^{j}+h_{k}^{i}\nabla_{m}\nabla_{n}R_{l}^{j}\Big)\nabla_{i}g_{j}.\label{ctr_3}
	\end{eqnarray}
Similarly, by plugging the explicit expression for $\Delta_2^{ij}$ (\ref{Deltaij_2}) into the first term in (\ref{I2_def}) and performing integrations by parts, we have
	\begin{eqnarray}
	 &  & \int d^{3}x\,2\sqrt{h}\Delta_{2}^{ij}\nabla_{i}g_{j}\nonumber \\
	 & \simeq & \int d^{3}x\bigg[-f\frac{\partial\Phi}{\partial(\nabla_{i_{1}}\nabla_{i_{2}}N)}2\Sigma_{ki_{1}i_{2}}^{lij}\nabla^{k}N\,\nabla_{l}\nabla_{i}g_{j}\nonumber \\
	 &  & -2f\frac{\partial\Phi}{\partial(\nabla_{i_{1}}\nabla_{i_{2}}\nabla_{i_{3}}N)}\nonumber \\
	 &  & \times\Big(\Sigma_{ki_{2}i_{3}}^{lij}\nabla^{k}N\,\nabla_{i_{1}}\nabla_{l}\nabla_{i}g_{j}+\mathcal{D}_{i_{1}i_{2}i_{3}}^{lij}\nabla_{l}\nabla_{i}g_{j}\Big)\nonumber \\
	 &  & -2f\frac{\partial\Phi}{\partial(\nabla_{i_{1}}\nabla_{i_{2}}\nabla_{i_{3}}\nabla_{i_{4}}N)}\Big(\Sigma_{ki_{3}i_{4}}^{lij}\nabla^{k}N\,\nabla_{i_{1}}\nabla_{i_{2}}\nabla_{l}\nabla_{i}g_{j}\nonumber \\
	 &  & +\mathcal{E}_{i_{1}i_{2}i_{3}i_{4}}^{l'lij}\nabla_{l'}\nabla_{l}\nabla_{i}g_{j}+\mathcal{F}_{i_{1}i_{2}i_{3}i_{4}}^{lij}\nabla_{l}\nabla_{i}g_{j}\Big)\bigg],\label{int_Delta_2}
	\end{eqnarray}
where tensors $\Sigma$, $\mathcal{D}$, $\mathcal{E}$ and $\mathcal{F}$ are defined in (\ref{Sigma_def}) and (\ref{Dcal})-(\ref{Fcal}), respectively.
Again, tedious calculations yield
	\begin{eqnarray}
	 &  & 2\Sigma_{ki_{1}i_{2}}^{lij}\nabla^{k}N\,\nabla_{l}\nabla_{i}g_{j}\nonumber \\
	 & = & \nabla_{i_{1}}\nabla_{i_{2}}\left(g^{k}\nabla_{k}N\right)-g^{k}\nabla_{k}\nabla_{i_{1}}\nabla_{i_{2}}N\nonumber \\
	 &  & -\left(h_{i_{1}}^{i}\nabla^{j}\nabla_{i_{2}}N+h_{i_{2}}^{i}\nabla_{i_{1}}\nabla^{j}N\right)\nabla_{i}g_{j},\label{ctr_4}
	\end{eqnarray}
	and
	\begin{eqnarray}
	 &  & 2\left(\Sigma_{ki_{2}i_{3}}^{lij}\nabla^{k}N\,\nabla_{i_{1}}\nabla_{l}\nabla_{i}g_{j}+\mathcal{D}_{i_{1}i_{2}i_{3}}^{lij}\nabla_{l}\nabla_{i}g_{j}\right)\nonumber \\
	 & = & \nabla_{i_{1}}\nabla_{i_{2}}\nabla_{i_{3}}\left(g^{k}\nabla_{k}N\right)-g^{k}\nabla_{k}\nabla_{i_{1}}\nabla_{i_{2}}\nabla_{i_{3}}N\nonumber \\
	 &  & -\Big(h_{i_{1}}^{i}\nabla^{j}\nabla_{i_{2}}\nabla_{i_{3}}N+h_{i_{2}}^{i}\nabla_{i_{1}}\nabla^{j}\nabla_{i_{3}}N\nonumber \\
	 &  & +h_{i_{3}}^{i}\nabla_{i_{1}}\nabla_{i_{2}}\nabla^{j}N\Big)\nabla_{i}g_{j},\label{ctr_5}
	\end{eqnarray}
	and
	\begin{eqnarray}
	 &  & 2\Big(\Sigma_{ki_{3}i_{4}}^{lij}\nabla^{k}N\,\nabla_{i_{1}}\nabla_{i_{2}}\nabla_{l}\nabla_{i}g_{j}\nonumber \\
	 &  & +\mathcal{E}_{i_{1}i_{2}i_{3}i_{4}}^{l'lij}\nabla_{l'}\nabla_{l}\nabla_{i}g_{j}+\mathcal{F}_{i_{1}i_{2}i_{3}i_{4}}^{lij}\nabla_{l}\nabla_{i}g_{j}\Big)\nonumber \\
	 & = & \nabla_{i_{1}}\nabla_{i_{2}}\nabla_{i_{3}}\nabla_{i_{4}}\left(g^{k}\nabla_{k}N\right)-g^{k}\nabla_{k}\nabla_{i_{1}}\nabla_{i_{2}}\nabla_{i_{3}}\nabla_{i_{4}}N\nonumber \\
	 &  & -\Big(h_{i_{1}}^{i}\nabla^{j}\nabla_{i_{2}}\nabla_{i_{3}}\nabla_{i_{4}}N+h_{i_{2}}^{i}\nabla_{i_{1}}\nabla^{j}\nabla_{i_{3}}\nabla_{i_{4}}N\nonumber \\
	 &  & +h_{i_{3}}^{i}\nabla_{i_{1}}\nabla_{i_{2}}\nabla^{j}\nabla_{i_{4}}N+h_{i_{4}}^{i}\nabla_{i_{1}}\nabla_{i_{2}}\nabla_{i_{3}}\nabla^{j}N\Big)\nabla_{i}g_{j},\qquad \label{ctr_6}
	\end{eqnarray}
In deriving (\ref{ctr_1})-(\ref{ctr_3}) and (\ref{ctr_4})-(\ref{ctr_6}), we frequently used the Bianchi identities as well as the definition of Riemann tensor as commutator of covariant derivatives.
Please note (\ref{ctr_1})-(\ref{ctr_3}) and (\ref{ctr_4})-(\ref{ctr_6}) are identities, in deriving which no integration by parts are performed.

Putting all the above together, we have 
	\begin{eqnarray}
	\mathcal{I}_{1} & = & \int d^{3}x\, f\mathcal{O}_{(1)}^{ij}\nabla_{i}g_{j},\label{I1_fin}\\
	\mathcal{I}_{2} & \equiv & \int d^{3}x\, f\bigg(\mathcal{O}_{(2)}^{ij}\nabla_{i}g_{j}-\sum_{n=0}^{4}\frac{\partial\Phi}{\partial\left(\nabla_{i_{1}}\cdots\nabla_{i_{n}}N\right)}\nonumber \\
	 &  & \times\nabla_{i_{1}}\cdots\nabla_{i_{n}}\left(g^{k}\nabla_{k}N\right)\bigg),\label{I2_fin}
	\end{eqnarray}
with
	\begin{eqnarray}
	 &  & \mathcal{O}_{(1)\phantom{i}j}^{\phantom{(1)}i}\nonumber \\
	 & \equiv & 2\frac{\partial\Phi}{\partial h_{ik}}h_{kj}-2h^{ik}\frac{\partial\Phi}{\partial h^{kj}}-2\pi^{ik}\frac{\partial\Phi}{\partial\pi^{jk}}\nonumber \\
	 &  & +\frac{\partial\Phi}{\partial R_{kl}}\left(h_{k}^{i}R_{jl}+h_{l}^{i}R_{kj}\right)\nonumber \\
	 &  & +\frac{\partial\Phi}{\partial\left(\nabla_{m}R_{kl}\right)}\left(h_{m}^{i}\nabla_{j}R_{kl}+h_{k}^{i}\nabla_{m}R_{jl}+h_{l}^{i}\nabla_{m}R_{kj}\right)\nonumber \\
	 &  & +\frac{\partial\Phi}{\partial\left(\nabla_{m}\nabla_{n}R_{kl}\right)}\Big(h_{m}^{i}\nabla_{j}\nabla_{n}R_{kl}+h_{n}^{i}\nabla_{m}\nabla_{j}R_{kl}\nonumber \\
	 &  & +h_{k}^{i}\nabla_{m}\nabla_{n}R_{jl}+h_{l}^{i}\nabla_{m}\nabla_{n}R_{kj}\Big).\label{Ocal_1}
	\end{eqnarray}
and
	\begin{eqnarray}
	 &  & \mathcal{O}_{(2)\phantom{i}j}^{\phantom{(2)}i}\nonumber \\
	 & \equiv & \frac{\partial\Phi}{\partial\left(\nabla_{i}N\right)}\nabla_{j}N\nonumber \\
	 &  & +\frac{\partial\Phi}{\partial\left(\nabla_{i_{1}}\nabla_{i_{2}}N\right)}\left(h_{i_{1}}^{i}\nabla_{j}\nabla_{i_{2}}N+h_{i_{2}}^{i}\nabla_{i_{1}}\nabla_{j}N\right)\nonumber \\
	 &  & +\frac{\partial\Phi}{\partial\left(\nabla_{i_{1}}\nabla_{i_{2}}\nabla_{i_{3}}N\right)}\Big(h_{i_{1}}^{i}\nabla_{j}\nabla_{i_{2}}\nabla_{i_{3}}N\nonumber \\
	 &  & +h_{i_{2}}^{i}\nabla_{i_{1}}\nabla_{j}\nabla_{i_{3}}N+h_{i_{3}}^{i}\nabla_{i_{1}}\nabla_{i_{2}}\nabla_{j}N\Big)\nonumber \\
	 &  & +\frac{\partial\Phi}{\partial\left(\nabla_{i_{1}}\nabla_{i_{2}}\nabla_{i_{3}}\nabla_{i_{4}}N\right)}\Big(h_{i_{1}}^{i}\nabla_{j}\nabla_{i_{2}}\nabla_{i_{3}}\nabla_{i_{4}}N\nonumber \\
	 &  & +h_{i_{2}}^{i}\nabla_{i_{1}}\nabla_{j}\nabla_{i_{3}}\nabla_{i_{4}}N+h_{i_{3}}^{i}\nabla_{i_{1}}\nabla_{i_{2}}\nabla_{j}\nabla_{i_{4}}N\nonumber \\
	 &  & +h_{i_{4}}^{i}\nabla_{i_{1}}\nabla_{i_{2}}\nabla_{i_{3}}\nabla_{j}N\Big),\label{Ocal_2}
	\end{eqnarray}
where again in (\ref{Ocal_2}) $n=0$ simply denotes $N$ without derivatives.
Finally, plugging (\ref{I1_fin})--(\ref{I2_fin}) into (\ref{PB_Phi_Ci_int}) and performing a further integration by parts in (\ref{I2_fin}) yield
	\begin{eqnarray}
	\left\{ \left\langle f,\Phi\right\rangle ,\left\langle g^{i},\mathcal{C}_{i}\right\rangle \right\} _{\mathrm{P}} & \simeq & -\int d^{3}x\bigg(\nabla_{i}f\,\Phi+\frac{\delta\left\langle f,\Phi\right\rangle }{\delta N}\nabla_{i}N\bigg)g^{i}\nonumber \\
	 &  & +\int d^{3}x\, f\left(\mathcal{O}_{(1)}^{ij}+\mathcal{O}_{(2)}^{ij}\right)\nabla_{i}g_{j},\label{PB_Phi_Ci_fun}
	\end{eqnarray}
where the spatial functional derivative of $\left\langle f,\Phi\right\rangle $ with respect to $N$ is given by
	\begin{eqnarray}
	\frac{\delta\left\langle f,\Phi\right\rangle }{\delta N} & = & \sqrt{h}\sum_{n=0}^{4}\left(-1\right)^{n}\nonumber \\
	 &  & \times\nabla_{i_{n}}\cdots\nabla_{i_{1}}\bigg(\frac{f}{\sqrt{h}}\frac{\partial\Phi}{\partial(\nabla_{i_{1}}\cdots\nabla_{i_{n}}N)}\bigg).\quad 
	\end{eqnarray}
While according to the identity (\ref{id_gen}), the second integral in (\ref{PB_Phi_Ci_fun}) exactly cancels out since (see Appendix \ref{sec:identity} for details)
	\begin{equation}
		\mathcal{O}^{ij}_{(1)} + \mathcal{O}^{ij}_{(2)} \equiv 0, \label{O1O2}
	\end{equation}
which implies
	\begin{equation}
	\left\{ \left\langle f,\Phi\right\rangle ,\left\langle g^{i},\mathcal{C}_{i}\right\rangle \right\} _{\mathrm{P}}=-\int d^{3}x\Big(\nabla_{i}f\,\Phi+\frac{\delta\left\langle f,\Phi\right\rangle }{\delta N}\nabla_{i}N\Big)g^{i}.\label{PB_Phi_Ci_fin}
	\end{equation}
By replacing $f\left(\vec{x}\right)=\int d^{3}x'\, f\left(\vec{x}'\right)\delta^{3}\left(\vec{x}-\vec{x}'\right)$
in (\ref{PB_Phi_Ci_fin}) and using the definition
	\begin{eqnarray}
	 &  & \left\{ \left\langle f,\Phi\right\rangle ,\left\langle g^{i},\mathcal{C}_{i}\right\rangle \right\} _{\mathrm{P}} \nonumber\\
	 & = & \int d^{3}x\int d^{3}x'\, f\left(\vec{x}'\right)g^{i}\left(\vec{x}\right)\left\{ \Phi\left(\vec{x}'\right),\mathcal{C}_{i}\left(\vec{x}\right)\right\} _{\mathrm{P}},
	\end{eqnarray}
one immediately arrives at (\ref{PB_Phi_Ci}). 
As we have mentioned, since (\ref{PB_Phi_Ci_fin}) and thus (\ref{PB_Phi_Ci}) are linear in $\Phi$, as long as $\Phi$ can be expressed in terms of a polynomial of $\pi^{ij}$ with each monomial taking the form (\ref{Phi_mono}),  (\ref{PB_Phi_Ci}) is valid.
This completes our proof.

We emphasize that the whole derivation of  (\ref{PB_Phi_Ci}) is only based on the assumption of $\Phi$ being a scalar density of unit weight defined on the phase space, which can be expressed in terms of a polynomial of $\pi^{ij}$ with coefficients being functions of $t$, $h_{ij}$, $N$, $R_{ij}$ and their spatial derivatives.
In particular, we never employed any concrete functional form for $\Phi$.
Moreover, although we truncate the spatial derivatives of $R_{ij}$ up to the second order and of $N$ up to the fourth order in order to present the explicit calculations, the same procedure can be generalized to include arbitrarily higher order spatial derivatives and  we expect (\ref{PB_Phi_Ci}) generally holds.

\section{Constraint algebra} \label{sec:cons_alg}

We are now ready to calculate the Poisson brackets among the constraints of our theory.

Among totally 10 types of Poisson brackets among the constraints (\ref{cons_all}), the following 6 of them are identically vanishing
	\begin{eqnarray}
	\left\{ \pi_{N}(\vec{x}),\pi_{N}(\vec{x}')\right\} _{\mathrm{P}}=0,\quad\left\{ \pi_{N}(\vec{x}),\pi_{i}(\vec{x}')\right\} _{\mathrm{P}} & = & 0,\quad\label{PB_zero1}\\
	\left\{ \pi_{i}(\vec{x}),\pi_{j}(\vec{x}')\right\} _{\mathrm{P}}=0,\quad\left\{ \pi_{N}(\vec{x}),\mathcal{C}_{i}(\vec{x}')\right\} _{\mathrm{P}} & = & 0,\quad\label{PB_zero2}\\
	\left\{ \pi_{i}(\vec{x}),\mathcal{C}(\vec{x}')\right\} _{\mathrm{P}}=0,\quad\left\{ \pi_{i}(\vec{x}),\mathcal{C}_{j}(\vec{x}')\right\} _{\mathrm{P}} & = & 0,\quad\label{PB_zero3}
	\end{eqnarray}
which can be checked easily by definition.
Exactly the same calculation in GR yields (see Appendix \ref{sec:PB_Ci_Cj} for a brief derivation)
	\begin{equation}
		\left\{ \left\langle f^{i},\mathcal{C}_{i}\right\rangle ,\left\langle g^{j},\mathcal{C}_{j}\right\rangle \right\} _{\mathrm{P}}=\big\langle \left(\pounds_{\bm{f}}\bm{g}\right)^{i},\mathcal{C}_{i}\big\rangle, \label{PB_Ci_Cj_fun}
	\end{equation}
with $\left(\pounds_{\bm{f}}\bm{g}\right)^{i}\equiv f^{j}\nabla_{j}g^{i}-g^{j}\nabla_{j}f^{i}$, which implies
	\begin{eqnarray}
	\left\{ \mathcal{C}_{i}\left(\vec{x}\right),\mathcal{C}_{j}\left(\vec{x}'\right)\right\} _{\mathrm{P}} & = & \mathcal{C}_{j}\left(\vec{x}\right)\nabla_{x^{i}}\delta^{3}\left(\vec{x}-\vec{x}'\right)\nonumber \\
	 &  & -\mathcal{C}_{i}\left(\vec{x}'\right)\nabla_{x'^{j}}\delta^{3}\left(\vec{x}-\vec{x}'\right).
	\end{eqnarray}
Generally, the Poisson bracket between $\mathcal{C}$ itself  $\left\{ \mathcal{C}\left(\vec{x}\right),\mathcal{C}\left(\vec{x}'\right)\right\}_{\mathrm{P}} $ does not respect the relation in GR: $\left\{ \mathcal{C}\left(\vec{x}\right),\mathcal{C}\left(\vec{x}'\right)\right\} _{\mathrm{P}}=\mathcal{C}^{i}\left(\vec{x}\right)\nabla_{x^{i}}\delta^{3}\left(\vec{x}-\vec{x}'\right)-\mathcal{C}^{i}\left(\vec{x}'\right)\nabla_{x'^{i}}\delta^{3}\left(\vec{x}-\vec{x}'\right)$, and thus does not vanish on the constraint surface.   
Its concrete expression is subject to the particular form of $\mathcal{C}$, which varies from model to model.
We neglect the calculation of $\left\{ \mathcal{C}\left(\vec{x}\right),\mathcal{C}\left(\vec{x}'\right)\right\}_{\mathrm{P}}$, which is irrelevant to our following analysis.

For our purpose, the nontrivial Poisson bracket is
	\begin{equation}
		\left\{ \pi_{N}\left(\vec{x}\right),\mathcal{C}\left(\vec{x}'\right)\right\} _{\mathrm{P}}=-\frac{\delta\mathcal{C}\left(\vec{x}'\right)}{\delta N\left(\vec{x}\right)} , \label{PB_piN_C}
	\end{equation}
where $\delta\mathcal{C}/\delta N$ is given by simply replacing $\Phi$ by $\mathcal{C}$ in (\ref{fd_Phi_N}):
	\begin{eqnarray}
	 &  & \frac{\delta\mathcal{C}(\vec{x}')}{\delta N(\vec{x})}\nonumber \\
	 & = & \delta^{3}\left(\vec{x}-\vec{x}'\right)\frac{\partial\mathcal{C}}{\partial N}+\sqrt{h}\sum_{n=1}\left(-1\right)^{n}\nonumber \\
	 &  & \times\nabla_{i_{n}}\cdots\nabla_{i_{1}}\bigg(\frac{\delta^{3}(\vec{x}-\vec{x}')}{\sqrt{h}}\frac{\partial\mathcal{C}}{\partial(\nabla_{i_{1}}\cdots\nabla_{i_{n}}N)}\bigg).\label{fd_C_N}
	\end{eqnarray}
On the other hand, from the analysis in Sec. \ref{sec:Ham_cons}, $\mathcal{C}$ can be written as a polynomial of $\pi^{ij}$ with coefficients being functions of $t$, $h_{ij}$, $N$, $R_{ij}$ and their spatial derivatives, and thus is a special case of the scalar density $\Phi$ analyzed in Sec. \ref{sec:PB_Phi_Ci}. Simply replacing $\Phi$ by $\mathcal{C}$ in (\ref{PB_Phi_Ci}) immediately yields
	\begin{equation}
		\left\{ \mathcal{C}(\vec{x}'),\mathcal{C}_{i}(\vec{x})\right\} _{\mathrm{P}}=-\mathcal{C}(\vec{x})\nabla_{i}\delta^{3}(\vec{x}-\vec{x}')-\frac{\delta\mathcal{C}(\vec{x}')}{\delta N(\vec{x})}\nabla_{i}N(\vec{x}), \label{PB_C_Ci}
	\end{equation}
where $\delta \mathcal{C}/ \delta N$ is also given by (\ref{fd_C_N}).

(\ref{PB_piN_C}) and (\ref{PB_C_Ci}) are the main results in this work.
In the case of GR, $\mathcal{C} = \partial ( N \tilde{\mathcal{C}} )/ \partial N \equiv \tilde{\mathcal{C}}$ does not depend on the lapse $N$, and thus all the Poisson brackets among the 8 constraints weakly vanish, which implies that all 8 constraints are first class. 
At this point, apparently there are infinite number of theories which are different from GR while satisfying $\frac{\delta\mathcal{C}\left(\vec{x}'\right)}{\delta N\left(\vec{x}\right)} = 0$. 
A subtle example is the ``non-projectable'' version of Ho\v{r}ava gravity \cite{Horava:2009uw}, in which the Lagrangian explicitly breaks general covariance while $N$ still serves as a Lagrange multiplier.
Such kind of theories, however, were found to be pathological \cite{Li:2009bg,Charmousis:2009tc,Koyama:2009hc}.
In the case of Ho\v{r}ava gravity, these pathologies were cured in \cite{Blas:2009qj} by adding invariants of acceleration $a_i = \partial_i \ln N$, such as $(a_i a^i)^n$ in the Lagrangian.
According to our analysis, it is clear that this is essentially to add nonlinear functional dependence on $N$ in the Hamiltonian, which prevents $N$ from being a Lagrange multiplier any more, and makes both $\pi_N\approx 0$ and $\mathcal{C}\approx 0$ to be second class.

In our case, as long as the constraint $\mathcal{C}$ has functional dependence on $N$, that is, at least one of the following derivatives
	\begin{equation}
		\frac{\partial\mathcal{C}}{\partial N},\qquad\frac{\partial\mathcal{C}}{\partial\left(\nabla_{i}N\right)},\qquad\frac{\partial\mathcal{C}}{\partial\left(\nabla_{i}\nabla_{j}N\right)},\qquad \cdots,
	\end{equation}
does not vanish identically on the constraint surface, we have
	\begin{equation}
		\frac{\delta\mathcal{C}\left(\vec{x}'\right)}{\delta N\left(\vec{x}\right)} \neq 0.
	\end{equation}
This happens when at least of one of $\mathcal{G}_{(n)}$'s and $\mathcal{V}$ in (\ref{L_gen}) depends on $N$ and/or its spatial derivatives.
In this case, on the constraint surface (i.e., in the sense of ``weak equality''), the matrix of Poisson brackets reads 
	\begin{center}
	\begin{tabular}{c|cccc}
	$\left\{ \cdot,\cdot\right\} _{\mathrm{P}}$ & $\pi_{N}(\vec{x}')$ & $\pi_{j}(\vec{x}')$ & $\mathcal{C}\left(\vec{x}'\right)$ & $\mathcal{C}_{j}\left(\vec{x}'\right)$\tabularnewline
	\hline 
	$\pi_{N}(\vec{x})$ & $0$ & $0$ & $-\frac{\delta\mathcal{C}(\vec{x}')}{\delta N\left(\vec{x}\right)}$ & $0$\tabularnewline
	$\pi_{i}\left(\vec{x}\right)$ & $0$ & $0$ & $0$ & $0$\tabularnewline
	$\mathcal{C}\left(\vec{x}\right)$ & $\frac{\delta\mathcal{C}\left(\vec{x}\right)}{\delta N\left(\vec{x}'\right)}$ & $0$ & $\left\{ \mathcal{C}(\vec{x}),\mathcal{C}(\vec{x}')\right\} $ & $-\frac{\delta\mathcal{C}\left(\vec{x}\right)}{\delta N\left(\vec{x}'\right)}\nabla_{x'^{i}}N(\vec{x}')$\tabularnewline
	$\mathcal{C}_{i}\left(\vec{x}\right)$ & $0$ & $0$ & $\frac{\delta\mathcal{C}(\vec{x}')}{\delta N(\vec{x})}\nabla_{i}N\left(\vec{x}\right)$ & $0$\tabularnewline
	\end{tabular}
	\par\end{center}
This is a $8\times 8$ matrix with 8 eigenvalues, of which six are identically zero, while two are non-vanishing\footnote{Indeed, for a $8\times 8$ antisymmetric with non-vanishing entries \[
\left(\begin{array}{cccc}
 &  & -x\\
 &  & 0_{i}\\
x & 0_{j} & 0 & -y_{j}\\
 &  & y_{i}
\end{array}\right),
\]
among totally 8 eigenvalues, 6 are identically vanishing, while the two non-zero eigenvalues are $\pm i\sqrt{x^{2}+y_{1}^{2}+y_{2}^{2}+y_{3}^{2}}$. It is also interesting to note this fact does not rely on the particular form of $y_i$'s.}.  
This fact implies that there are always 8 linearly independent combinations of the 8 constraints in (\ref{cons_all}), of which 6 are first class and 2 are second class.

At this point, note (\ref{PB_C_Ci}) implies $\mathcal{C}_i \approx 0$ themselves are not first-class\footnote{This was also pointed out in the Hamiltonian analysis \cite{Lin:2014jga} for  the model in \cite{Gleyzes:2014dya}.}. 
Nevertheless, it is easy to show that
	\begin{equation}
		\left\{ \mathcal{C}\left(\vec{x}'\right),\pi_{N}\left(\vec{x}\right)\nabla_{i}N\left(\vec{x}\right)\right\} _{\mathrm{P}}=\frac{\delta\mathcal{C}\left(\vec{x}'\right)}{\delta N\left(\vec{x}\right)}\nabla_{i}N\left(\vec{x}\right),
	\end{equation}
which exactly reproduces the second term in (\ref{PB_C_Ci}).
Thus we may introduce a ``shifted'' momentum constraint as the linear combination of $\mathcal{C}_i$ and $\pi_N$: 
	\begin{equation}
		\tilde{\mathcal{C}}_i \equiv \mathcal{C}_i + \pi_N \nabla_i N, \label{Ci_tilde}
	\end{equation}	
which yields
	\begin{equation}
		\big\{ \mathcal{C}\left(\vec{x}\right),\tilde{\mathcal{C}}_{i}\left(\vec{x}'\right)\big\} _{\mathrm{P}}=\mathcal{C}\left(\vec{x}'\right)\nabla_{x^{i}}\delta^{3}\left(\vec{x}-\vec{x}'\right)\approx0.
	\end{equation}
Remarkably, although our theory can be very general, this ``shifted'' momentum constraint $\tilde{\mathcal{C}}_i$ is the same one as introduced in \cite{Lin:2014jga,Gleyzes:2014qga}.
It is also straightforward to verify that
	\begin{eqnarray}
	\big\{\pi_{N}(\vec{x}),\tilde{\mathcal{C}}_{i}(\vec{x}')\big\}_{\mathrm{P}} & = & \nabla_{x^{i}}\delta^{3}(\vec{x}-\vec{x}')\pi_{N}(\vec{x}')\approx0,\label{PB_piN_Ci_tilde}\\
	\big\{\pi_{i}(\vec{x}),\tilde{\mathcal{C}}_{j}(\vec{x}')\big\}_{\mathrm{P}} & = & 0,\label{PB_pii_Ci_tilde}\\
	\big\{\tilde{\mathcal{C}}_{i}(\vec{x}),\tilde{\mathcal{C}}_{j}(\vec{x}')\big\}_{\mathrm{P}} & = & \tilde{\mathcal{C}}_{j}(\vec{x})\nabla_{x^{i}}\delta^{3}(\vec{x}-\vec{x}')\nonumber \\
	 &  & -\tilde{\mathcal{C}}_{i}(\vec{x}')\nabla_{x'^{j}}\delta^{3}(\vec{x}-\vec{x}')\approx0.\label{PB_Ci_Cj_tilde}
	\end{eqnarray}
Thus, in the new set of 8 linearly independent constraints
	\begin{equation}
		\pi_N\approx 0, \qquad \pi_i\approx 0,\qquad \mathcal{C}\approx 0, \qquad \tilde{\mathcal C}_i\approx 0, \label{cons_all_2}
	\end{equation}
$\pi_{i}$ and $\tilde{\mathcal C}_i$ are six first class constraints, and $\pi_N$ and $\mathcal{C}$ are two second class constraints since $\left\{ \pi_{N},\mathcal{C}\right\} _{\mathrm{P}}\not\approx 0$.

As a consistency check, it is important to verify that the algebra is closed, i.e., no further secondary constraint is generated. 
In fact, straightforward manipulations yield (see Appendix \ref{sec:cons_Ci} for a derivation)
	\begin{eqnarray}
	\frac{\mathrm{d}}{\mathrm{d}t} \mathcal{C}_{i}(\vec{x}) & \approx & \left\{ \mathcal{C}_{i}\left(\vec{x}\right),H_{\mathrm{E}}\right\} _{\mathrm{P}}\nonumber \\
	 & = & \nabla_{i}N(\vec{x})\,\mathcal{C}(\vec{x})+\nabla_{i}N^{j}(\vec{x})\mathcal{C}_{j}(\vec{x})\nonumber \\
	 &  & -\int d^{3}x'\, N^{j}(\vec{x}')\nabla_{x'^{j}}\delta^{3}(\vec{x}-\vec{x}')\,\mathcal{C}_{i}(\vec{x}')\nonumber \\
	 &  & +\nabla_{i}N(\vec{x})\frac{\delta\left\langle \lambda_{\mathcal{C}},\mathcal{C}\right\rangle }{\delta N(\vec{x})},\label{td_Ci}
	\end{eqnarray}
which implies $\mathcal{C}_i$ is not automatically preserved if $\mathcal{C}$ has functional dependence on $N$. Nevertheless, we have
	\begin{eqnarray}
	\frac{\mathrm{d} }{\mathrm{d}t}(\pi_{N}\nabla_{i}N) & \approx & \left\{ \pi_{N}\nabla_{i}N ,H_{\mathrm{E}}\right\} _{\mathrm{P}}\nonumber \\
	 & = & \nabla_{i}\lambda^{N}\pi_{N}-\nabla_{i}N\,\mathcal{C}-\nabla_{i}N\frac{\delta\langle\lambda_{\mathcal{C}},\mathcal{C}\rangle}{\delta N},\quad \label{td_piNdN}
	\end{eqnarray}
which implies that the combination $\tilde{\mathcal{C}}_i$ defined in (\ref{Ci_tilde}) is  preserved on the constraint surface since
	\begin{eqnarray}
	\frac{\mathrm{d}}{\mathrm{d}t}\tilde{\mathcal{C}}_{i} & \equiv & \frac{\mathrm{d}}{\mathrm{d}t}\left(\mathcal{C}_{i}+\pi_{N}\nabla_{i}N\right)\nonumber \\
	 & \approx & \nabla_{i}N\frac{\delta\left\langle \lambda_{\mathcal{C}},\mathcal{C}\right\rangle }{\delta N}-\nabla_{i}N\frac{\delta\left\langle \lambda_{\mathcal{C}},\mathcal{C}\right\rangle }{\delta N}=0.
	\end{eqnarray}
It is also easy to show that $\mathrm{d}\pi_{i}/\mathrm{d}t \approx \left\{ \pi_{i},H_{\mathrm{E}}\right\} \approx0$.
On the other hand, the condition
	\begin{equation}
		\frac{\mathrm{d}}{\mathrm{d}t}\mathcal{C} \approx  \frac{\partial \mathcal{C}}{\partial t}+\left\{ \mathcal{C},H_{\mathrm{E}}\right\} _{\mathrm{P}} \approx 0,
	\end{equation}
together with $\mathrm{d}\pi_N/\mathrm{d}t\approx 0$ simply fix the Lagrange multiplies $\lambda_{\mathcal{C}}$ and $\lambda^N$ instead of generating new constraint, since $\mathcal{C}\approx 0$ and $\pi_N\approx 0$ are second class. To conclude, the 8 constraints in (\ref{cons_all}) or equivalently in (\ref{cons_all_2}) are the all constraints in our theory.

According to the usual counting degrees of freedom for the constraint systems, each first class constraint together with the associated gauge fixing condition eliminate two canonical variables, while each second class constraint eliminates one canonical variable. 
The number of independent physical degrees of freedom in our theory (\ref{L_gen}) is thus given by
	\begin{eqnarray}
	\text{number of d.o.f.} & = & \frac{1}{2}\big(2\times\text{number of canonical variables}\nonumber \\
	 &  & -2\times\text{number of first class constraints}\nonumber \\
	 &  & -\text{number of second class constraints}\big)\nonumber \\
	 & = & \frac{1}{2}\left(2\times10-2\times6-2\right)\nonumber \\
	 & = & 3.
	\end{eqnarray}

\section{Conclusion}

Recently, there is an increasing interest in exploring scalar tensor theories ``beyond Horndeski'', which propagate the correct number of degrees of freedom while having higher order equations of motion.
When being written in the unitary gauge, such kind of theories correspond to a class of gravity theories respecting only spatial diffeomorphism.
In this work, we have performed a detailed Hamiltonian constraint analysis of a class of such spatially covariant gravity theories proposed in \cite{Gao:2014soa}, of which the Lagrangian is given by (\ref{L_gen}).
With a very general setup, we have shown that as long as the lapse function $N$ enters the Hamiltonian nonlinearly, both the primary and secondary constraints associated with $N$ become second class.
As a result, besides the two degrees of freedom of the usual transverse and traceless tensor gravitons as in GR, our theory propagates an additional scalar mode, which can be viewed as the longitudinal  graviton, at the fully nonlinear level.

By construction, the Lagrangian (\ref{L_gen}) includes the model proposed recently in \cite{Gleyzes:2014dya} as a special case, of which similar Hamiltonian analysis was also performed in \cite{Lin:2014jga} and \cite{Gleyzes:2014qga}.
Spatial derivatives of $N$ or $R_{ij}$ were not included in the model \cite{Gleyzes:2014dya}, which are generally allowed in (\ref{L_gen}).
Moreover, the analysis in \cite{Lin:2014jga,Gleyzes:2014qga} only considered specific Lagrangians quadratic in the extrinsic curvature, which correspond to the case of $\Pi^{ij}[K]$ defined in (\ref{Piij_def}) being linear in $K_{ij}$. 
Our analysis, on the other hand, is based on a very general setup.
In particular, our analysis does not rely on any concrete functional form for the Lagrangian.
We only assume the Hamiltonian can be formally expressed as a polynomial of $\pi^{ij}$, with coefficients being general functions of $t$, $h_{ij}$, $N$ and $R_{ij}$ and their spatial derivatives, which is a natural result within our general framework.
Although we have included spatial derivatives of $R_{ij}$ up to the second order, and of $N$ up to the fourth order in order to make explicit calculations, we expect the same procedures in this work can be extended to the cases with higher order spatial derivatives and the conclusion will not change.


\acknowledgments

I would like to thank
Kazuya Koyama,
Shinji Mukohyama,
Gianmassimo Tasinato,
Masahide Yamaguchi
for useful discussions and comments.
I am grateful to Nathalie Deruelle for support and to AstroParticule et Cosmologie (APC) in Paris for hospitality, during my visit in which this work was initiated.
I also wish to thank the Yukawa Institute for Theoretical Physics (YITP) at Kyoto University for hospitality, during my visit in which this work was finalized.
I was supported by JSPS Grant-in-Aid for Scientific Research No. 25287054.

\appendix
\section{Functional derivatives}

\subsection{$\mathcal{C}_{i} $} \label{sec:fd_Ci}

The calculation is the same as in GR. Here we collect the steps for completeness. By definition
	\begin{eqnarray}
	\left\langle f^{i},\mathcal{C}_{i}\right\rangle  & \equiv & -\int d^{3}x\,2\sqrt{h}h_{ik}f^{k}\nabla_{j}\left(\frac{1}{\sqrt{h}}\pi^{ij}\right)\nonumber \\
	 & \simeq & \int d^{3}x\,2h_{ik}\nabla_{j}f^{k}\pi^{ij},
	\end{eqnarray}
varying which with respect to $h_{ij}$ and $\pi^{ij}$ yields
	\begin{eqnarray}
	\delta\left\langle f^{i},\mathcal{C}_{i}\right\rangle  & = & \int d^{3}x\,2\Big[\delta h_{ij}\nabla_{k}f^{j}\pi^{ik}\nonumber \\
	 &  & +h_{ij}\delta\left(\nabla_{k}f^{i}\right)\pi^{jk}+h_{ik}\nabla_{j}f^{k}\delta\pi^{ij}\Big]\nonumber \\
	 & = & \int d^{3}x\,2\Big[\delta h_{ij}\nabla_{k}f^{j}\pi^{ik}\nonumber \\
	 &  & +h_{ij}\delta\Gamma_{kl}^{i}f^{l}\pi^{jk}+\nabla_{(i}f_{j)}\delta\pi^{ij}\Big].
	\end{eqnarray}
Using
	\[
		\delta\Gamma_{kl}^{i}=\frac{1}{2}h^{ij}\left(\nabla_{k}\delta h_{lj}+\nabla_{l}\delta h_{jk}-\nabla_{j}\delta h_{kl}\right),
	\]
we have
	\begin{eqnarray}
	 &  & \delta\left\langle f^{i},\mathcal{C}_{i}\right\rangle \nonumber \\
	 & = & \int d^{3}x\,2\Big(\delta h_{ij}\nabla_{k}f^{j}\pi^{ik}+\frac{1}{2}\nabla_{l}\delta h_{jk}f^{l}\pi^{jk}+\nabla_{(i}f_{j)}\delta\pi^{ij}\Big)\nonumber \\
	 & \simeq & \int d^{3}x\,2\Big[\delta h_{ij}\nabla_{k}f^{j}\pi^{ik}-\frac{1}{2}\sqrt{h}\nabla_{k}\bigg(\frac{1}{\sqrt{h}}f^{k}\pi^{ij}\bigg)\delta h_{ij}\nonumber \\
	 &  & +\nabla_{(i}f_{j)}\delta\pi^{ij}\Big],
	\end{eqnarray}
which implies
	\begin{eqnarray}
	\frac{\delta\left\langle f^{k},\mathcal{C}_{k}\right\rangle }{\delta h_{ij}} & = & 2\nabla_{k}f^{(i}\pi^{j)k}-\sqrt{h}\nabla_{k}\left(f^{k}\frac{\pi^{ij}}{\sqrt{h}}\right),\label{fd_Ci_h}\\
	\frac{\delta\left\langle f^{k},\mathcal{C}_{k}\right\rangle }{\delta\pi^{ij}} & = & 2\nabla_{(i}f_{j)}.\label{fd_Ci_pi}
	\end{eqnarray}
(\ref{fd_Ci_h}) and (\ref{fd_Ci_pi}) are exactly the same as in GR.
	
\subsection{$\Phi$} \label{sec:fd_Phi}

We focus on the monomial
	\begin{equation}
		\Phi=\frac{1}{h^{\frac{n-1}{2}}}\Gamma_{i_{1}j_{1},\cdots,i_{n}j_{n}}\pi^{i_{1}j_{1}}\cdots\pi^{i_{n}j_{n}}, \label{sd_ansatz}
	\end{equation}
where $\Gamma_{i_{1}j_{1},\cdots,i_{n}j_{n}}$ is a tensorial function of $t$, $h_{ij}$, $N$ and $R_{ij}$ as well as their spatial derivatives.

The variation of $\Phi$ with respect to $\pi^{ij}$ is simply given in (\ref{fd_Phi_pi}). To evaluate the variation with respect to $h_{ij}$ is much involved. 
First we have
	\begin{eqnarray}
	\delta_{h}\left\langle f,\Phi\right\rangle  & = & \int d^{3}x\, f\bigg(-\frac{n-1}{2}\frac{1}{h^{\frac{n+1}{2}}}hh^{ij}\delta h_{ij}\Gamma_{i_{1}j_{1},\cdots,i_{n}j_{n}}\nonumber \\
	 &  & +\frac{1}{h^{\frac{n-1}{2}}}\delta_{h}\Gamma_{i_{1}j_{1},\cdots,i_{n}j_{n}}\bigg)\pi^{i_{1}j_{1}}\cdots\pi^{i_{n}j_{n}}\nonumber \\
	 & \equiv & \int d^{3}x\,\bigg(-\frac{n-1}{2}f\Phi h^{ij}\delta h_{ij}\nonumber \\
	 &  & +f\frac{1}{h^{\frac{n-1}{2}}}\delta_{h}\Gamma_{i_{1}j_{1},\cdots,i_{n}j_{n}}\pi^{i_{1}j_{1}}\cdots\pi^{i_{n}j_{n}}\bigg),\label{var_h_fun}
	\end{eqnarray}
where in the second line in (\ref{var_h_fun}), we used the definition of $\Phi$ (\ref{sd_ansatz}).
Generally, $\Gamma_{i_{1}j_{1},\cdots,i_{n}j_{n}}$ may contain arbitrarily
higher orders of spatial derivatives of $R_{ij}$ and $N$. In order
to evaluate $\delta_{h}\Gamma_{i_{1}j_{1},\cdots,i_{n}j_{n}}$ definitely, in
this following, we restrict the spatial derivatives of $R_{ij}$ in
$\Gamma_{i_{1}j_{1},\cdots,i_{n}j_{n}}$ up to the second order, and of $N$ up to  the fourth order. Precisely, we consider $\Gamma_{i_{1}j_{1},\cdots,i_{n}j_{n}}$ to be tensorial function of
	\begin{eqnarray*}
	 &  & h_{ij},\quad h^{ij},\\
	 &  & R_{kl},\quad\nabla_{m}R_{kl},\quad\nabla_{m}\nabla_{n}R_{kl},\\
	 &  & N,\quad\nabla_{i}N,\quad\nabla_{i}\nabla_{j}N,\quad\nabla_{i}\nabla_{j}\nabla_{k}N,\quad\nabla_{i}\nabla_{j}\nabla_{k}\nabla_{l}N,
	\end{eqnarray*}
as well as time $t$.
This has already included wide class of models and is sufficient to show the logic and the generality of our proof.
The variation with respect to $h_{ij}$ thus yields
	\begin{eqnarray}
	 &  & \delta_{h}\Gamma_{i_{1}j_{1},\cdots,i_{n}j_{n}}\nonumber \\
	 & = & \frac{\partial\Gamma_{i_{1}j_{1},\cdots,i_{n}j_{n}}}{\partial h_{ij}}\delta h_{ij}+\frac{\partial\Gamma_{i_{1}j_{1},\cdots,i_{n}j_{n}}}{\partial h^{ij}}\delta h^{ij}\nonumber \\
	 &  & +\frac{\partial\Gamma_{i_{1}j_{1},\cdots,i_{n}j_{n}}}{\partial R_{ij}}\delta R_{ij}+\frac{\partial\Gamma_{i_{1}j_{1},\cdots,i_{n}j_{n}}}{\partial(\nabla_{k}R_{ij})}\delta(\nabla_{k}R_{ij})\nonumber \\
	 &  & +\frac{\partial\Gamma_{i_{1}j_{1},\cdots,i_{n}j_{n}}}{\partial(\nabla_{k}\nabla_{l}R_{ij})}\delta(\nabla_{k}\nabla_{l}R_{ij})+\frac{\partial\Gamma_{i_{1}j_{1},\cdots,i_{n}j_{n}}}{\partial(\nabla_{i}\nabla_{j}N)}\delta(\nabla_{i}\nabla_{j}N)\nonumber \\
	 &  & +\frac{\partial\Gamma_{i_{1}j_{1},\cdots,i_{n}j_{n}}}{\partial(\nabla_{i}\nabla_{j}\nabla_{k}N)}\delta(\nabla_{i}\nabla_{j}\nabla_{k}N)\nonumber \\
	 &  & +\frac{\partial\Gamma_{i_{1}j_{1},\cdots,i_{n}j_{n}}}{\partial(\nabla_{i}\nabla_{j}\nabla_{k}\nabla_{l}N)}\delta(\nabla_{i}\nabla_{j}\nabla_{k}\nabla_{l}N).\label{var_h_gen}
	\end{eqnarray}
Note the variation of derivatives of $N$ starts from the second derivatives $\nabla_{i}\nabla_{j}N$, since $\nabla_{i}N\equiv \partial_i N$ has nothing to do with the metric.

The linear variations of $R_{ij}$ and its derivatives with respect to the metric are given by
	\begin{equation}
	\delta R_{kl}=\mathcal{A}_{kl}^{ijl_{1}l_{2}}\nabla_{l_{1}}\nabla_{l_{2}}\delta h_{ij},\label{rt_var}
	\end{equation}
	\begin{equation}
	\delta(\nabla_{m}R_{kl})=\mathcal{A}_{kl}^{ijl_{1}l_{2}}\nabla_{m}\nabla_{l_{1}}\nabla_{l_{2}}\delta h_{ij}-\mathcal{B}_{mkl}^{ijl'}\nabla_{l'}\delta h_{ij},\label{d1rt_var}
	\end{equation}
and
	\begin{eqnarray}
	 \delta(\nabla_{m}\nabla_{n}R_{kl}) & = & \mathcal{A}_{kl}^{ijl_{1}l_{2}}\nabla_{m}\nabla_{n}\nabla_{l_{1}}\nabla_{l_{2}}\delta h_{ij}\nonumber \\
	 &  & -\mathcal{B}_{nkl}^{ijl'}\nabla_{m}\nabla_{l'}\delta h_{ij}-\mathcal{C}_{mnkl}^{ijl'}\nabla_{l'}\delta h_{ij},\qquad \label{d2rt_var}
	\end{eqnarray}
with
	\begin{eqnarray}
	\mathcal{A}_{kl}^{ijl_{1}l_{2}} & \equiv & h^{l_{1}k'}\Sigma_{k'kl}^{l_{2}ij}-h^{k_{1}k_{2}}\Sigma_{k_{1}k_{2}(k}^{l_{2}ij}h_{l)}^{l_{1}},\label{Acal}\\
	\mathcal{B}_{mkl}^{ijl'} & \equiv & R_{k}^{k'}\Sigma_{k'lm}^{l'ij}+R_{l}^{k'}\Sigma_{k'km}^{l'ij},\label{Bcal}\\
	\mathcal{C}_{mnkl}^{ijl'} & \equiv & \nabla^{k'}R_{kl}\Sigma_{k'mn}^{l'ij}+\nabla_{m}R_{k}^{k'}\Sigma_{k'ln}^{l'ij}\nonumber \\
	 &  & +\nabla_{m}R_{l}^{k'}\Sigma_{k'kn}^{l'ij}+\nabla_{n}R_{k}^{k'}\Sigma_{k'lm}^{l'ij}\nonumber \\
	 &  & +\nabla_{n}R_{l}^{k'}\Sigma_{k'km}^{l'ij},\label{Ccal}
	\end{eqnarray}
where  $\Sigma_{nkl}^{mij}$ is defined by
	\begin{equation}
	\Sigma_{nkl}^{mij}\equiv\frac{1}{2}\left(h_{k}^{m}h_{l}^{(i}h_{n}^{j)}+h_{l}^{m}h_{n}^{(i}h_{k}^{j)}-h_{n}^{m}h_{k}^{(i}h_{l}^{j)}\right).\label{Sigma_def}
	\end{equation}
For the linear variations of $\nabla_{i}\nabla_{j}N$ etc., we have
	\begin{equation}
	\delta(\nabla_{i_{1}}\nabla_{i_{2}}N)=-\Sigma_{ki_{1}i_{2}}^{lij}\nabla^{k}N\,\nabla_{l}\delta h_{ij},\label{d2N_var}
	\end{equation}
	\begin{equation}
	\delta(\nabla_{i_{1}}\nabla_{i_{2}}\nabla_{i_{3}}N)=-\Sigma_{ki_{2}i_{3}}^{lij}\nabla^{k}N\,\nabla_{i_{1}}\nabla_{l}\delta h_{ij}-\mathcal{D}_{i_{1}i_{2}i_{3}}^{lij}\nabla_{l}\delta h_{ij},\label{d3N_var}
	\end{equation}
	and
	\begin{eqnarray}
	 &  & \delta(\nabla_{i_{1}}\nabla_{i_{2}}\nabla_{i_{3}}\nabla_{i_{4}}N)\nonumber \\
	 & = & -\Sigma_{ki_{3}i_{4}}^{lij}\nabla^{k}N\,\nabla_{i_{1}}\nabla_{i_{2}}\nabla_{l}\delta h_{ij}\nonumber \\
	 &  & -\mathcal{E}_{i_{1}i_{2}i_{3}i_{4}}^{l'lij}\nabla_{l'}\nabla_{l}\delta h_{ij}-\mathcal{F}_{i_{1}i_{2}i_{3}i_{4}}^{lij}\nabla_{l}\delta h_{ij},\label{d4N_var}
	\end{eqnarray}
with
	\begin{eqnarray}
	\mathcal{D}_{i_{1}i_{2}i_{3}}^{lij} & \equiv & \Sigma_{ki_{2}i_{3}}^{lij}\nabla_{i_{1}}\nabla^{k}N+\Sigma_{ki_{1}i_{2}}^{lij}\nabla^{k}\nabla_{i_{3}}N\nonumber \\
	 &  & +\Sigma_{ki_{1}i_{3}}^{lij}\nabla_{i_{2}}\nabla^{k}N,\label{Dcal}
	\end{eqnarray}
	\begin{align}
	\mathcal{E}_{i_{1}i_{2}i_{3}i_{4}}^{l'lij}= & h_{i_{1}}^{l'}\Sigma_{ki_{3}i_{4}}^{lij}\nabla_{i_{2}}\nabla^{k}N+h_{i_{2}}^{l'}\Sigma_{ki_{3}i_{4}}^{lij}\nabla_{i_{1}}\nabla^{k}N\nonumber \\
	 & +h_{i_{1}}^{l'}\Sigma_{ki_{2}i_{3}}^{lij}\nabla^{k}\nabla_{i_{4}}N+h_{i_{1}}^{l'}\Sigma_{ki_{2}i_{4}}^{lij}\nabla_{i_{3}}\nabla^{k}N,\label{Ecal}
	\end{align}
and
	\begin{align}
	\mathcal{F}_{i_{1}i_{2}i_{3}i_{4}}^{lij}= & \Sigma_{ki_{3}i_{4}}^{lij}\nabla_{i_{1}}\nabla_{i_{2}}\nabla^{k}N+\Sigma_{ki_{2}i_{3}}^{lij}\nabla_{i_{1}}\nabla^{k}\nabla_{i_{4}}N\nonumber \\
	 & +\Sigma_{ki_{2}i_{4}}^{lij}\nabla_{i_{1}}\nabla_{i_{3}}\nabla^{k}N+\Sigma_{ki_{1}i_{2}}^{lij}\nabla^{k}\nabla_{i_{3}}\nabla_{i_{4}}N\nonumber \\
	 & +\Sigma_{ki_{1}i_{3}}^{lij}\nabla_{i_{2}}\nabla^{k}\nabla_{i_{4}}N+\Sigma_{ki_{1}i_{4}}^{lij}\nabla_{i_{2}}\nabla_{i_{3}}\nabla^{k}N,\label{Fcal}
	\end{align}	
where $\Sigma_{nkl}^{mij}$ is the same as defined in (\ref{Sigma_def}).

Plugging (\ref{var_h_gen}), (\ref{rt_var})-(\ref{d2rt_var}) and (\ref{d2N_var})-(\ref{d4N_var}) into (\ref{var_h_fun}), using $\delta h^{ij}=-h^{ik}h^{jl}\delta h_{kl}$ and integrating by parts the derivatives of $\delta h_{ij}$, finally we arrive at
	\begin{equation}
		\frac{\delta\left\langle f,\Phi\right\rangle }{\delta h_{ij}}=-\frac{n-1}{2}f\,\Phi\, h^{ij}+\sqrt{h}\left(\Delta_{1}^{ij}+\Delta_{2}^{ij}\right), \label{fd_Phi_h}
	\end{equation}
with
	\begin{align}
	& \Delta_{1}^{ij}\nonumber\\
	= & \frac{1}{\sqrt{h}}f\left(\frac{\partial\Phi}{\partial h_{ij}}-h^{ik}h^{jl}\frac{\partial\Phi}{\partial h^{kl}}\right)\nonumber \\
	 & +\nabla_{l_{2}}\nabla_{l_{1}}\left(f\frac{1}{\sqrt{h}}\frac{\partial\Phi}{\partial R_{kl}}\mathcal{A}_{kl}^{ijl_{1}l_{2}}\right)\nonumber \\
	 & +\nabla_{l'}\left(f\frac{1}{\sqrt{h}}\frac{\partial\Phi}{\partial\left(\nabla_{m}R_{kl}\right)}\mathcal{B}_{mkl}^{ijl'}\right)\nonumber \\
	 & -\nabla_{l_{2}}\nabla_{l_{1}}\nabla_{m}\left(f\frac{1}{\sqrt{h}}\frac{\partial\Phi}{\partial\left(\nabla_{m}R_{kl}\right)}\mathcal{A}_{kl}^{ijl_{1}l_{2}}\right)\nonumber \\
	 & +\nabla_{l'}\left(f\frac{1}{\sqrt{h}}\frac{\partial\Phi}{\partial\left(\nabla_{m}\nabla_{n}R_{kl}\right)}\mathcal{C}_{mnkl}^{ijl'}\right)\nonumber \\
	 & -\nabla_{l'}\nabla_{m}\left(f\frac{1}{\sqrt{h}}\frac{\partial\Phi}{\partial\left(\nabla_{m}\nabla_{n}R_{kl}\right)}\mathcal{B}_{nkl}^{ijl'}\right)\nonumber \\
	 & +\nabla_{l_{2}}\nabla_{l_{1}}\nabla_{n}\nabla_{m}\left(f\frac{1}{\sqrt{h}}\frac{\partial\Phi}{\partial\left(\nabla_{m}\nabla_{n}R_{kl}\right)}\mathcal{A}_{kl}^{ijl_{1}l_{2}}\right),\label{Deltaij_1}
	\end{align}
	and
	\begin{align}
	 & \Delta_{2}^{ij}\nonumber \\
	= & \nabla_{l}\left(\frac{1}{\sqrt{h}}f\frac{\partial\Phi}{\partial\left(\nabla_{i_{1}}\nabla_{i_{2}}N\right)}\Sigma_{ki_{1}i_{2}}^{lij}\nabla^{k}N\right)\nonumber \\
	 & -\nabla_{l}\nabla_{i_{1}}\left(\frac{1}{\sqrt{h}}f\frac{\partial\Phi}{\partial\left(\nabla_{i_{1}}\nabla_{i_{2}}\nabla_{i_{3}}N\right)}\Sigma_{ki_{2}i_{3}}^{lij}\nabla^{k}N\right)\nonumber \\
	 & +\nabla_{l}\left(\frac{1}{\sqrt{h}}f\frac{\partial\Phi}{\partial\left(\nabla_{i_{1}}\nabla_{i_{2}}\nabla_{i_{3}}N\right)}\mathcal{D}_{i_{1}i_{2}i_{3}}^{lij}\right)\nonumber \\
	 & +\nabla_{l}\nabla_{i_{2}}\nabla_{i_{1}}\left(\frac{1}{\sqrt{h}}f\frac{\partial\Phi}{\partial\left(\nabla_{i_{1}}\nabla_{i_{2}}\nabla_{i_{3}}\nabla_{i_{4}}N\right)}\Sigma_{ki_{3}i_{4}}^{lij}\nabla^{k}N\right)\nonumber \\
	 & -\nabla_{l}\nabla_{l'}\left(\frac{1}{\sqrt{h}}f\frac{\partial\Phi}{\partial\left(\nabla_{i_{1}}\nabla_{i_{2}}\nabla_{i_{3}}\nabla_{i_{4}}N\right)}\mathcal{E}_{i_{1}i_{2}i_{3}i_{4}}^{l'lij}\right)\nonumber \\
	 & +\nabla_{l}\left(\frac{1}{\sqrt{h}}f\frac{\partial\Phi}{\partial\left(\nabla_{i_{1}}\nabla_{i_{2}}\nabla_{i_{3}}\nabla_{i_{4}}N\right)}\mathcal{F}_{i_{1}i_{2}i_{3}i_{4}}^{lij}\right),\label{Deltaij_2}
	\end{align}
where $\mathcal{A}_{kl}^{ijl_{1}l_{2}}$, $\mathcal{B}_{mkl}^{ijl'}$ etc are given in (\ref{Acal})-(\ref{Ccal}) and (\ref{Dcal})-(\ref{Fcal}).
Please note in deriving (\ref{fd_Phi_h})-(\ref{Deltaij_2}), we never assume any concrete functional form for $\Phi$ in (\ref{sd_ansatz}).
Moreover, at this point it is not necessary to evaluate the covariant derivatives in (\ref{Deltaij_1}) and (\ref{Deltaij_2}) explicitly, since which will be removed by integrations by parts again when calculating the Poisson brackets.

Note we also have
	\begin{eqnarray*}
	 &  & \delta_{N}\left\langle f,\Phi\right\rangle \\
	 & = & \int d^{3}x\, f\left(\vec{x}\right)\Bigg(\frac{\partial\Phi}{\partial N}\delta N+\frac{\partial\Phi}{\partial\left(\nabla_{i}N\right)}\delta\left(\nabla_{i}N\right)\\
	 &  & +\frac{\partial\Phi}{\partial\left(\nabla_{i}\nabla_{j}N\right)}\delta\left(\nabla_{i}\nabla_{j}N\right)+\cdots\Bigg)\\
	 & \simeq & \int d^{3}x\Bigg[f\frac{\partial\Phi}{\partial N}\delta N-\sqrt{h}\nabla_{i}\left(f\frac{1}{\sqrt{h}}\frac{\partial\Phi}{\partial\left(\nabla_{i}N\right)}\right)\delta N\\
	 &  & +\sqrt{h}\nabla_{j}\nabla_{i}\left(f\frac{1}{\sqrt{h}}\frac{\partial\Phi}{\partial\left(\nabla_{i}\nabla_{j}N\right)}\right)\delta N+\cdots\Bigg],
	\end{eqnarray*}
which implies
	\begin{eqnarray}
	\frac{\delta\left\langle f,\Phi\right\rangle }{\delta N} & = & f\frac{\partial\Phi}{\partial N}+\sqrt{h}\sum_{n=1}\left(-1\right)^{n}\nonumber \\
	 &  & \times\nabla_{i_{n}}\cdots\nabla_{i_{1}}\Bigg(\frac{f}{\sqrt{h}}\frac{\partial\Phi}{\partial(\nabla_{i_{1}}\cdots\nabla_{i_{n}}N)}\Bigg),\qquad \label{fd_Phi_N_fun}
	\end{eqnarray}
for an arbitrary function $\Phi$ on the phase space.


\section{Poisson bracket $\left\{ \mathcal{C}_{i}\left(\vec{x}\right),\mathcal{C}_{j}\left(\vec{x}'\right)\right\} _{\mathrm{P}}$} \label{sec:PB_Ci_Cj}

Here we briefly review the derivation of Poisson bracket $\left\{ \mathcal{C}_{i}\left(\vec{x}\right),\mathcal{C}_{j}\left(\vec{x}'\right)\right\} _{\mathrm{P}}$ for completeness, which is exactly the same as in GR.
By definition and plugging (\ref{fd_Ci_h})-(\ref{fd_Ci_pi}),
	\begin{eqnarray}
	 &  & \left\{ \left\langle f^{i},\mathcal{C}_{i}\right\rangle ,\left\langle g^{j},\mathcal{C}_{j}\right\rangle \right\} _{\mathrm{P}}\nonumber \\
	 & = & \int d^{3}x\Bigg(\frac{\delta\left\langle f^{i},\mathcal{C}_{i}\right\rangle }{\delta h_{kl}\left(\vec{x}\right)}\frac{\delta\left\langle g^{j},\mathcal{C}_{j}\right\rangle }{\delta\pi^{kl}\left(\vec{x}\right)}-\frac{\delta\left\langle f^{i},\mathcal{C}_{i}\right\rangle }{\delta\pi^{kl}\left(\vec{x}\right)}\frac{\delta\left\langle g^{j},\mathcal{C}_{j}\right\rangle }{\delta h_{kl}\left(\vec{x}\right)}\Bigg)\nonumber \\
	 & = & \int d^{3}x\bigg[2\nabla_{i}f^{k}\pi^{il}\nabla_{k}g_{l}-2\sqrt{h}\nabla_{i}\left(f^{i}\frac{\pi^{kl}}{\sqrt{h}}\right)\nabla_{k}g_{l}\nonumber \\
	 &  & -2\nabla_{i}g^{k}\pi^{il}\nabla_{k}f_{l}+2\sqrt{h}\nabla_{i}\left(g^{i}\frac{\pi^{kl}}{\sqrt{h}}\right)\nabla_{k}f_{l}\bigg].\label{PB_Ci_Cj_t1}
	\end{eqnarray}
Integrating by parts $\nabla_i$ in each term in (\ref{PB_Ci_Cj_t1}) and then using the definition of $\mathcal{C}_i$ yield
	\begin{eqnarray}
	 &  & \left\{ \left\langle f^{i},\mathcal{C}_{i}\right\rangle ,\left\langle g^{j},\mathcal{C}_{j}\right\rangle \right\} _{\mathrm{P}}\nonumber \\
	 & \simeq & \int d^{3}y\Big[\mathcal{C}^{l}\left(f^{k}\nabla_{k}g_{l}-g^{k}\nabla_{k}f_{l}\right)\nonumber \\
	 &  & -2f^{k}\,\pi^{il}\left[\nabla_{i},\nabla_{k}\right]g_{l}+2g^{k}\,\pi^{il}\left[\nabla_{i},\nabla_{k}\right]f_{l}\Big],\label{PB_Ci_Cj_t2}
	\end{eqnarray}
where the last two terms in (\ref{PB_Ci_Cj_t2}) drop out since
	\begin{eqnarray*}
	 &  & -2f^{k}\pi^{il}\left[\nabla_{i},\nabla_{k}\right]g_{l}+2g^{k}\pi^{il}\left[\nabla_{i},\nabla_{k}\right]f_{l}\\
	 & = & 2\left(-R_{iklm}+R_{imlk}\right)f^{k}g^{m}\pi^{il}\equiv0.
	\end{eqnarray*}

\section{A mathematical identity of derivatives} \label{sec:identity}

For any scalar type (index-free) object $\mathcal{I}$ made by arbitrary contraction of a set of tensors with lower indices $\left\{ T_{l_{1}\cdots l_{N}}\right\} $ (including the metric $h_{ij}$) and a set of tensors with upper indices $\left\{ S^{k_{1}\cdots k_{M}}\right\} $ (including the inverse metric $h^{ij}$), we have the following equality
	\begin{eqnarray}
	 &  & \sum_{\left\{ S^{k_{1}\cdots k_{M}}\right\} }\frac{\partial\mathcal{I}}{\partial S^{k_{1}\cdots k_{M}}}\sum_{m=1}^{M}h_{j}^{k_{m}}S^{k_{1}\cdots k_{m-1}ik_{m+1}\cdots k_{M}}\nonumber \\
	 & = & \sum_{\left\{ T_{l_{1}\cdots l_{N}}\right\} }\frac{\partial\mathcal{I}}{\partial T_{l_{1}\cdots l_{N}}}\sum_{n=1}^{N}h_{l_{n}}^{i}T_{l_{1}\cdots l_{n-1}jl_{n+1}\cdots l_{N}},\label{id_gen}
	\end{eqnarray}
where $\sum_{\left\{ S^{k_{1}\cdots k_{M}}\right\} }$ and $\sum_{\left\{ T_{l_{1}\cdots l_{N}}\right\} }$ denote summation over all tensors.
For example, for a scalar type function $\mathcal{I}$ contracted by the following tensors (tensors do not necessarily respect any symmetry) 
	\begin{equation}
		\mathcal{I}=\mathcal{I}\left(U^{i},S^{ij},A^{ijk},X^{ijkl},V_{i},T_{ij},B_{ijk},Y_{ijkl}\right) ,
	\end{equation}
we have
	\begin{align}
	 & \frac{\partial\mathcal{I}}{\partial U^{j}}U^{i}+\frac{\partial\mathcal{I}}{\partial S^{kl}}\Big(h_{j}^{k}S^{il}+h_{j}^{l}S^{ki}\Big)\nonumber \\
	 & +\frac{\partial\mathcal{I}}{\partial A^{mkl}}\Big(h_{j}^{m}A^{ikl}+h_{j}^{k}A^{mil}+h_{j}^{l}A^{mki}\Big)\nonumber \\
	 & +\frac{\partial\mathcal{I}}{\partial X^{mnkl}}\Big(h_{j}^{m}X^{inkl}+h_{j}^{n}X^{mikl}+h_{j}^{k}X^{mnil}+h_{j}^{l}X^{mnki}\Big)\nonumber \\
	= & \frac{\partial\mathcal{I}}{\partial V_{i}}V_{j}+\frac{\partial\mathcal{I}}{\partial T_{kl}}\Big(h_{k}^{i}T_{jl}+h_{l}^{i}T_{kj}\Big)\nonumber \\
	 & +\frac{\partial\mathcal{I}}{\partial B_{mkl}}\Big(h_{m}^{i}B_{jkl}+h_{k}^{i}B_{mjl}+h_{l}^{i}B_{mkj}\Big)\nonumber \\
	 & +\frac{\partial\mathcal{I}}{\partial Y_{mnkl}}\Big(h_{m}^{i}Y_{jnkl}+h_{n}^{i}Y_{mjkl}+h_{k}^{i}Y_{mnjl}+h_{l}^{i}Y_{mnkj}\Big).\label{id_gen_4}
	\end{align}
Please keep in mind that derivatives with respect to the metric and its inverse must also be included.

In our case, (\ref{O1O2}) is simply a special case of the general identity (\ref{id_gen}). In fact, by noticing that the monomial (\ref{Phi_mono}) is composed of the following tensors with indices: 
	\begin{eqnarray*}
	\text{upper:}\quad &  & h^{ij},\quad\pi^{ij},\\
	\text{lower:}\quad &  & h_{ij},\quad R_{ij},\quad\nabla_{k}R_{ij},\quad\nabla_{k}\nabla_{l}R_{ij},\\
	 &  & \nabla_{i}N,\quad\nabla_{i}\nabla_{j}N,\quad\nabla_{i}\nabla_{j}\nabla_{k}N,\quad\nabla_{i}\nabla_{j}\nabla_{k}\nabla_{l}N,
	\end{eqnarray*}
inserting the derivatives of $\Phi$ with respect to these tensors into (\ref{id_gen}), we immediately get (\ref{O1O2}).

\section{Conservation of $\mathcal{C}_i$ and $\pi_N \nabla_i N$} \label{sec:cons_Ci}

First we take time derivative of $\left\langle f^{i},\mathcal{C}_{i}\right\rangle $, which is given by
	\begin{eqnarray}
	 &  & \frac{\mathrm{d}}{\mathrm{d}t}\left\langle f^{i},\mathcal{C}_{i}\right\rangle \nonumber \\
	 & \approx & \left\{ \left\langle f^{i},\mathcal{C}_{i}\right\rangle ,H_{\mathrm{E}}\right\} _{\mathrm{P}}\nonumber \\
	 & = & \big\{\left\langle f^{i},\mathcal{C}_{i}\right\rangle ,\langle1,N\tilde{\mathcal{C}}\rangle\big\}_{\mathrm{P}}+\left\{ \left\langle f^{i},\mathcal{C}_{i}\right\rangle ,\left\langle N^{j},\mathcal{C}_{j}\right\rangle \right\} _{\mathrm{P}}\nonumber \\
	 &  & +\left\{ \left\langle f^{i},\mathcal{C}_{i}\right\rangle ,\left\langle \lambda_{\mathcal{C}},\mathcal{C}\right\rangle \right\} _{\mathrm{P}},\label{td_fiCi_ori}
	\end{eqnarray}
where $H_{\mathrm{E}}$ is the extended Hamiltonian given in (\ref{H_ex}).
For the first term in (\ref{td_fiCi_ori}), formally replacing $f\rightarrow 1$ in (\ref{PB_Phi_Ci_fin}) yields
	\[
		\left\{ \left\langle 1,\Phi\right\rangle ,\left\langle g^{i},\mathcal{C}_{i}\right\rangle \right\} _{\mathrm{P}}=-\int d^{3}x\frac{\delta\left\langle 1,\Phi\right\rangle }{\delta N}g^{i}\nabla_{i}N,
	\]
which implies
	\begin{eqnarray}
	\left\{ \left\langle f^{i},\mathcal{C}_{i}\right\rangle ,\big\langle 1,N\tilde{\mathcal{C}}\big\rangle \right\} _{\mathrm{P}} & = & \int d^{3}x\frac{\delta\big\langle 1,N\tilde{\mathcal{C}}\big\rangle }{\delta N}f^{i}\nabla_{i}N\nonumber \\
	 & \equiv & \int d^{3}x\,\mathcal{C}f^{i}\nabla_{i}N\approx0,
	\end{eqnarray}
where we used the definition  $\mathcal{C}\equiv\frac{\delta\left\langle 1,N\tilde{\mathcal{C}}\right\rangle }{\delta N}$.
For the second term in (\ref{td_fiCi_ori}), simply replacing $g^i\rightarrow N^i$ in  (\ref{PB_Ci_Cj_fun}) yields\footnote{This is justified when $f^i$ does not depend on phase space variables. }
	\begin{equation}
		\left\{ \left\langle f^{i},\mathcal{C}_{i}\right\rangle ,\left\langle N^{j},\mathcal{C}_{j}\right\rangle \right\} _{\mathrm{P}}=\int d^{3}x\,\left(\pounds_{\bm{f}}\bm{N}\right)^{i}\mathcal{C}_{i}\approx 0.
	\end{equation}
Thus finally we have $\frac{\mathrm{d}}{\mathrm{d}t}\left\langle f^{i},\mathcal{C}_{i}\right\rangle \approx \left\langle f^{i}\nabla_{i}N,\frac{\delta\left\langle \lambda_{\mathcal{C}},\mathcal{C}\right\rangle }{\delta N}\right\rangle $.

For the time evolution of $\pi_{N}\nabla_{i}N$, straightforward calculation shows
	\begin{equation}
	\begin{aligned} & \frac{\mathrm{d}}{\mathrm{d}t}\left\langle f^{i},\pi_{N}\nabla_{i}N\right\rangle \\
	\approx & \left\{ \left\langle f^{i},\pi_{N}\nabla_{i}N\right\rangle ,H_{\mathrm{E}}\right\} _{\mathrm{P}}\\
	= & \int d^{3}x\bigg(\frac{\delta\langle f^{i},\pi_{N}\nabla_{i}N\rangle}{\delta N}\frac{\delta H_{\mathrm{E}}}{\delta\pi_{N}}-\frac{\delta\langle f^{i},\pi_{N}\nabla_{i}N\rangle}{\delta\pi_{N}}\frac{\delta H_{\mathrm{E}}}{\delta N}\bigg)\\
	= & \int d^{3}x\bigg(-\nabla_{i}(f^{i}\pi_{N})\lambda^{N}-f^{i}\nabla_{i}N\frac{\delta\langle1,N\tilde{\mathcal{C}}\rangle}{\delta N}\\
	 & -f^{i}\nabla_{i}N\frac{\delta\left\langle \lambda_{\mathcal{C}},\mathcal{C}\right\rangle }{\delta N}\bigg)\\
	\simeq & \left\langle f^{i}\nabla_{i}\lambda^{N},\pi_{N}\right\rangle -\left\langle f^{i}\nabla_{i}N,\mathcal{C}\right\rangle -\left\langle f^{i}\nabla_{i}N,\frac{\delta\left\langle \lambda_{\mathcal{C}},\mathcal{C}\right\rangle }{\delta N\left(\vec{x}\right)}\right\rangle ,
	\end{aligned}
	\end{equation}
that is $\frac{\mathrm{d}}{\mathrm{d}t}\left\langle f^{i},\pi_{N}\nabla_{i}N\right\rangle \approx-\left\langle f^{i}\nabla_{i}N,\frac{\delta\left\langle \lambda_{\mathcal{C}},\mathcal{C}\right\rangle }{\delta N\left(\vec{x}\right)}\right\rangle $.


\bibliography{Gao}

\end{document}